\documentclass[11pt,preprin,showpacs,preprintnumbers,eqsecnum,amsmath,amssymb,nofootinbib,times,natbib]{revtex4}
\usepackage{eurosym}
\usepackage[utf8]{inputenc}
\pdfoutput=1
\usepackage{graphicx}
\usepackage{color}
\usepackage{braket}
\usepackage{dcolumn}
\usepackage{bm,url}
\usepackage[linktocpage]{hyperref}
\usepackage{subfigure}
\usepackage{amsfonts}
\usepackage[usenames,dvipsnames,svgnames]{xcolor}  
\usepackage{hyperref}   
\definecolor{oxfordblue}{rgb}{0.0, 0.13, 0.28}
\definecolor{burgundy}{rgb}{0.5, 0.0, 0.13}
\definecolor{darkolivegreen}{rgb}{0.33, 0.42, 0.18}
\definecolor{darkblue}{rgb}{0,0,0.5}
\definecolor{richcarmine}{rgb}{0.84, 0.0, 0.25}
\definecolor{darkblue}{rgb}{0,0,0.5}
\definecolor{bluer}{rgb}{0.00,0.50,0.75}{}
\hypersetup{colorlinks=true, citecolor=green, linkcolor=bluer,
urlcolor = bluer, filecolor=magenta}
\begin{document}
\newcommand\be{\begin{equation}}
\newcommand\ee{\end{equation}}
\newcommand\bea{\begin{eqnarray}}
\newcommand\eea{\end{eqnarray}}
\newcommand\bseq{\begin{subequations}} 
\newcommand\eseq{\end{subequations}}
\newcommand\bcas{\begin{cases}}
\newcommand\ecas{\end{cases}}
\newcommand{\p}{\partial}
\newcommand{\f}{\frac}

\title{  Black Hole Superradiance in the Presence of Lorentz Symmetry Violation}

\author {\textbf{Mohsen Khodadi}}
\email{m.khodadi@ipm.ir}
\affiliation{\small  \textit{School of Astronomy, Institute for Research in Fundamental Sciences (IPM)\\
		P. O. Box 19395-5531, Tehran, Iran}}

\date{\today}

\begin{abstract}
In this paper  we consider the massive scalar perturbation on the top of a small spinning-like black hole in context of Einstein-bumblebee modified gravity in order to probe 	 the role of spontaneous Lorentz symmetry breaking on the superradiance scattering and corresponding instability.
We show that at the low-frequency limit of the scalar wave the superradiance scattering will be enhanced with the Lorentz-violating parameter $\alpha<0$ and will be weakened with $\alpha>0$. Moreover, by addressing the black hole bomb issue, we extract an improved bound in the instability regime indicating that $\alpha<0$ increases the parameter space of the scalar field instability, while $\alpha>0$ decreases it.

\end{abstract}
\maketitle

\section{Introduction} \label{Int}
Thanks to recent observations stellar-mass black hole collisions by ground-based gravitational-wave detectors \cite{Abbott:2016blz,TheLIGOScientific:2016pea, LIGOScientific:2018mvr} and capturing the first images of the supermassive black hole's shadow at the center of galaxy M87* by the Event Horizon Telescope (EHT) \cite{Akiyama:2019cqa,Akiyama:2019eap}, lots of attentions from astronomers around the world have been attracted to reconsider the physics of black hole more deeply.  Newly, also has discovered a stellar-mass black hole which is closer to our Solar System than any other found to date so that by forming part of a triple system it surprisingly can be seen with the naked eye \cite{Rivinius_2020}.
It is known among astrophysicists that the supermassive black holes are the most likely driving
power of spectacular jets in high energetic sources such as active galactic nuclei (AGN)\footnote{A new study \cite{Piotrovich:2020kae} claims that AGNs actually maybe wormhole mouths rather than supermassive black holes. } and quasars at the center of the many galaxies \cite{Peterson:2000sw,Emami:2019mzi}. However, it should be noted that black holes in nature are not an isolated system rather prone to destroy due to some external perturbations. Therefore, it is well-motivated to stability analysis of black holes exposed to matter fields in high energy astrophysics.
In this regard, one of the mysterious phenomena widely used in the domain of modern physics, is named \emph{``superradiance''}.
Generally speaking, we expect this phenomenon reveals in astrophysical black holes belong to the Kerr-Newman family
with three characteristic parameters mass, charge and spin.
Black hole superradiance actually is an amplification of scattered test field off the rotating and/or charged black hole's horizon which
due to energy extracting from background then the reflection coefficient of the wave equation solution to exceed the unit,
see leading papers \cite{Zel1}-\cite{Detweiler:1980uk}\footnote{If we want to be precise, we have to say that the first treatment
of superradiance though wrongly, comes from the seminal work of Klein \cite{Klein:1929zz} in the context of non-relativistic
quantum mechanics.}.  This phenomenon in a sense categorized into black hole perturbation issues so that it occurs just for the bosonic fields perturbations
(scalar, electromagnetic and gravitational waves) and not for the massive or massless fermionic test fields
such as neutrinos, \cite{Unruh:1973bda,Chandrasekhar:1976ap,Kim:1997hy}. An impressive message
that black hole superradiance has to us is that in the curved spacetime, one can expect to energy extraction from
the vacuum via a quite classic process, without regard to any quantum considerations.
This is due to this fact that the existence of an event horizon
leads to the emergence of an essential ingredient for this phenomenon, namely
an intrinsic dissipative mechanism in vacuum \cite{Brito:2015oca}.

Concerning the Kerr black hole, the energy extraction results in decreasing whether two characteristic parameters mass and spin without violating the second law of black hole known as area law \footnote{Note that black hole's area law is valid just for black holes perturbed
by matter fields satisfying the weak energy condition. For this reason, one usually does not expect this law
works for the black holes subject to the fermionic field perturbations.}
\cite{Bekenstein:1973mi}. As a consequence, using direct relation between black hole area
and entropy \cite{Bekenstein:1973ur}, it was found that in superradiance phenomenon, we just
deal with rotational energy extraction from the black hole, not information.
So, this phenomena has not any conflict with the event horizon's concept
as a one-way membrane which means that there is not information transfer from the interior region  to the exterior region of a black hole. There is also another classical
mechanism for extracting energy from Kerr black hole known as \emph{``Penrose process''}. It actually was realized earlier than black hole superradiance through Penrose's Gedankenexperiment in which a mass of the matter decomposed into ergoregion so that one of which obtains negative energy, while the other escapes to infinity with more energy than before decomposition\footnote{To have a detailed study of these two energy extraction processes, we refer the interested reader to \cite{Vicente:2018mxl} and also review paper \cite{Brito:2015oca}} \cite{Pen}.
However, it was understood that this process phenomenologically can not be a favor energy extraction mechanism in the sens that it requires an accurate
timed breakup of the incoming relativistic particles \cite{Bard}.

Because of transferring energy from black hole to bosonic test field, superradiance phenomenon can generate some instabilities in the background  due to traping the superradiant modes near the black hole.
More precisely, these instabilities leads to a new
phenomenon calling \emph{``black hole bomb''} because of growing the trapping superradiant modes  exponentially between black hole and a turning point.
There are some studies in both the contexts of charged and rotating black holes indicating that the turning point can be created generally
in two ways.
First, it can sourceed by massive scalar field or asymptotically Anti-de-site spacetime, second due to placing a reflective surface like a mirror or cavity around the black hole, see \cite{Cardoso:2004hs}-\cite{Li:2019tns} and references therein. Concerning the second case, may raise the inquiry in mind that is it applicable in realistic situations? The ionized matters such as plasma or accretion disc around astrophysical black holes due to their ability in reflecting of low-frequency electromagnetic waves,
are prone to playing the role of the mirror \cite{Brito:2015oca}.
In some cases the superradiant instabilities may lead to emerging exotic black hole solutions with additional parameters violating the no-hair hypothesis \footnote{In recent years, there have been extensive attempts to
test the validity of the no-hair hypothesis via  confronting its variety statements with some observations, see
\cite{Johannsen:2016uoh,Isi:2019aib,Khodadi:2020jij} for instance.}, \cite{Herdeiro:2015tia}-\cite{Rahmani:2020vvv}.

Although  during the last century GR have passed many tests with  great success and  it has been admitted
as the standard theory of gravity, still there are some doors to make alternative theories to GR to consider physical phenomena at extremely and small scales\footnote{In \cite{Salvatore} one can find more discussions about different motivations of going beyond Einstein gravity.}.
As an important factor from a theoretical viewpoint, having an ultraviolet completion
of GR is severely favorable.
From  observational point of view,
GR is not able to describe some gravitational phenomena at the large scale such as the dark side of the universe and
seems that something fundamental is still missing.

Black holes can be used a useful  probe
in the strong regime of gravity in order to check  possible high-energy modifications to GR. The imprint of these modifications can leave some signatures in astrophysical phenomena such
as strong gravity effects happening in the vicinity celestial bodies like astrophysical black holes and neutron
stars.
So it is of fundamental importance to investigate astrophysical attributes of black hole in alternative theories of gravity. Given that the astrophysical phenomenon under our attention \emph{i.e.} black hole
superradiance highly is sensitive to underlying geometries, hence in recent years, we have witnessed many
types of research about it and its side issues within the extended framework of modified gravity theories,
see \cite{Pani:2011gy}-\cite{Liu:2020lwc}.

In continuing this, we are interested in studying quench or favor superradiance of the spinning black holes in an extended Lorentz-violating gravity known as the ``Einstein-bumblebee model'' (EBM) \cite{Kostelecky:2003fs}. EBM, contains the innovative idea  \emph{``spontaneous Lorentz symmetry breaking''} (LSB) which actually
comes from one of the popular theoretical frameworks to solve quantum gravity issue \emph{i.e.} string theory.
It is worth mentioning to note that the  Lorentz symmetry as the fundamental underlying symmetry of  two successful
field theories describing the universe \emph{i.e.} GR and the particle standard model, may break at quantum gravity scales.
The LSB's idea, has resulted in emergence of an effective field theory known as \emph{``standard model extension''} (SME)
which via it the particle standard model plus GR and every operator may break the Lorentz symmetry, will bring together in one framework \cite{Kostelecky:1995qk}-\cite{Coleman:1998ti}. SME let one for further look for the LSB in multiple frameworks
such as: high energy particle physics and astrophysics. SME can be used in most modern analyses of experimental results.
EBM is actually one of simple models belong to
LSB in which the Lorentz symmetry dynamically breaks by an axial vector field
known as \emph{``bumblebee field''}. In other words, the presence of LSB in a local Lorentz frame usually is diagnosed through
a nonzero vacuum value for one or more quantities carrying local Lorentz indices which specifically in EBM it is signaled via
the bumblebee field.
In recent years, we see a remarkable enthusiasm among people for study of interesting physical themes in the framework of EBM,
see as example \cite{Seifert:2009gi}-\cite{Chen:2020qyp}, along with references therein.
However, in the meantime, can be seen a vacancy of studying the black hole superradiance.
This research is potentially valuable in the sense that due to going to the quantum gravity realm via black holes then
it is possible to take the contribution of LSB into superradiance phenomenon.

Aiming to probe the role of LSB on the black hole superradiance phenomenon (in particular amplification factor and superradiance
frequency parameters region)  as well as the instability related to it, we have organized this paper as follows.
In section \ref{sec:background},
we provide a short review of the derivation of the exact Kerr-like black hole solution in the EBM, based on \cite{Ding:2019mal}.
In section \ref{sec:swb},
by restoring to a semi-analytically method we release the superradiance amplification factor of bumblebee Kerr-like black hole subject
to a massive scalar perturbation. Subsequently we move to investigation on the superradiant instability of underlying system in section \ref{sec:SI}.
Finally, we close our paper with a summarize of concluding discussions and results in section \ref{sec:Con}.
Across this work, we use the natural units $c = G = \hbar = 1$.
\section{Exact Kerr-like black hole solution in Einstein-bumblebee model}\label{sec:background}
Here, without going to into details we will have a reviwe of the EBM-based Kerr-like black hole solution
which recently released in \cite{Ding:2019mal}.
Bumblebee theory of gravity as a subclass
of aether models in which Lorentz symmetry is spontaneously broken due to a nonzero vacuum expectation
value of the bumblebee vector field $B_{\mu}$ and a proper choosing of potential as well, generally describe
by the action \cite{Kostelecky:2003fs}
\begin{eqnarray}\label{action}
S_{EB}=
\int d^4x\sqrt{-g}\Bigg(\frac{1}{2\kappa}(R+\chi B^{\mu}B^{\nu}R_{\mu\nu})-\frac{1}{4}B^{\mu\nu}B_{\mu\nu}
-V(B^{\mu}B_{\mu}\pm b^2)\Bigg)~,~~~~\kappa\equiv8\pi
\end{eqnarray}
where $\chi$ and $b^2$, are respectively the real coupling and numerical constants so that
the former is indeed responsible for controlling the non-minimal coupling between the Ricci curvature
$R_{\mu\nu}$ and bumblebee field $B_\mu$. In the above action, $B_{\mu\nu}$ is named bumblebee field strength
and enjoys $U(1)$ gauge invariant, i.e. $B_{\mu\nu}=\partial_{\mu}B_{\nu}-\partial_{\nu}B_{\mu}$. Actually, in
this setup the Lorentz symmetry is broken via breaking of the $U(1)$ symmetry of the bumblebee field potential
i.e. the collapse of $V$ into a non-zero minimum at $B^{\mu}B_{\mu}\pm b^2=0$ and $V'(b_{\mu}b^{\mu})=0$, with
$b^{\mu}=\langle B^{\mu}\rangle$ which has constant magnitude $b_{\mu}b^{\mu}=\mp b^2$.
Note that here the prime symbol refers to the differentiation with respect to $b_{\mu}b^{\mu}$.

Concerning the action (\ref{action}), the gravitational field equation in vacuum takes the following form
\begin{align}\label{einstein}
R_{\mu\nu}=\kappa T_{\mu\nu}^B+2\kappa g_{\mu\nu}V
+\frac{1}{2}\kappa g_{\mu\nu} B^{\alpha\beta}B_{\alpha\beta}-
\kappa g_{\mu\nu} B^{\alpha}B_{\alpha}V'+\frac{\chi}{4}g_{\mu\nu}\nabla^2(B^{\alpha}B_{\alpha})
+\frac{\chi}{2}g_{\mu\nu}\nabla_{\alpha}\nabla_{\beta}(B^{\alpha}B^{\beta})~,
\end{align}
with the bumblebee energy momentum tensor $T_{\mu\nu}^B$ as
\begin{eqnarray}
&&T_{\mu\nu}^B=-B_{\mu\alpha}B^{\alpha}_{\;\nu}-\frac{1}{4}g_{\mu\nu} B^{\alpha\beta}B_{\alpha\beta}- g_{\mu\nu}V+
2B_{\mu}B_{\nu}V'\nonumber\\
&&+\frac{\chi}{\kappa}\Bigg(\frac{1}{2}g_{\mu\nu}B^{\alpha}B^{\beta}R_{\alpha\beta}
-B_{\mu}B^{\alpha}R_{\alpha\nu}-B_{\nu}B^{\alpha}R_{\alpha\mu}\nonumber\\
&&+\frac{1}{2}\nabla_{\alpha}\nabla_{\mu}(B^{\alpha}B_{\nu})
+\frac{1}{2}\nabla_{\alpha}\nabla_{\nu}(B^{\alpha}B_{\mu})
-\frac{1}{2}\nabla^2(B^{\mu}B_{\nu})-\frac{1}{2}
g_{\mu\nu}\nabla_{\alpha}\nabla_{\beta}(B^{\alpha}B^{\beta})\Bigg)~.
\end{eqnarray}
In case of fixing the bumblebee field to $B_\mu=b_\mu$ and subsequently $V=0=V'$, then
the extended gravitational field equation (\ref{einstein}), finally reduces to
\begin{eqnarray}\label{bar}
&&R_{\mu\nu}+\kappa b_{\mu\alpha}b^{\alpha}_{\;\nu}-\frac{\kappa}{4}g_{\mu\nu}
b^{\alpha\beta}b_{\alpha\beta}+\chi b_{\mu}b^{\alpha}R_{\alpha\nu}
+\chi b_{\nu}b^{\alpha}R_{\alpha\mu}
-\frac{\chi}{2}g_{\mu\nu}b^{\alpha}b^{\beta}R_{\alpha\beta}-\nonumber\\
&&\frac{\chi}{2}\Bigg(
\nabla_{\alpha}\nabla_{\mu}(b^{\alpha}b_{\nu})
+\nabla_{\alpha}\nabla_{\nu}(b^{\alpha}b_{\mu})
-\nabla^2(b_{\mu}b_{\nu})\Bigg)=0~.
\end{eqnarray}

By adopting the standard Boyer-Lindquist coordinates, the underlying gravity model admits a
Kerr-like black hole solution as \cite{Ding:2019mal}
\begin{eqnarray}\label{metric0}
ds^2=g_{tt}dt^2+g_{rr}dr^2+g_{\theta\theta}d\theta^2+g_{\phi\phi}
d\phi^2+2g_{t\phi}dtd\phi~,
\end{eqnarray}
where
\begin{eqnarray}
&&g_{tt}=- \Big(1-\frac{2Mr}{\varrho^2}\Big),\;\;\;\;g_{rr}=\frac{\varrho^2}{\Delta_r},\;\;
g_{\theta\theta}=\varrho^2,\;\;g_{\phi\phi}=\frac{A\sin^2\theta}{\varrho^2}
\nonumber\\
&&g_{t\phi}=-\frac{2Mra\sqrt{\alpha+1}\sin^2\theta}{\varrho^2}~,
\end{eqnarray}
with
\begin{align}
\varrho^2=r^2+(\alpha+1)a^2\cos^2\theta,\;\;\;
\Delta_\alpha=\frac{r^2-2Mr}{\alpha+1}+a^2,\;\;A=\bigg(r^2+(\alpha+1)a^2\bigg)^2-\Delta_\alpha(\alpha+1)^2 a^2\sin^2\theta~.
\end{align} In the above solution the spontaneously Lorentz symmetry breaking, addressed by the constant parameter $\alpha$,
so that by discarding it then this solution is reduced to the standard Kerr black hole.
More exactly,
the metric (\ref{metric0}) denotes a purely radial Lorentz-violating black hole solution with the ADM mass $M$, the
angular momentum (per unit mass) $a$ and also a Lorentz-violating parameter $\alpha$.
The location of the inner and outer horizons
acquire via
\begin{eqnarray}
r_{\mp}=M\mp\sqrt{M^2-a^2(\alpha+1)}~,
\end{eqnarray} which are correspond to the Cushy and event horizons labeled with $r_{ch}$ and $r_{eh}$, respectively.
At first glance, one may tell the radius of inner and outer horizons are not distinguishable from
their standard counterpart since we can easily absorb the Lorentz-violating parameter $\alpha$ via redefying $a\rightarrow
\hat{a}=\sqrt{\alpha+1}~a$ ($\alpha>-1$).
Although this seems to be true, it may carry the misleading message that the metric
(\ref{metric0}) is nothing but a standard Kerr metric. However, after applying a such redefinition for $a$, the metric (\ref{metric0})
re-express as
\begin{eqnarray}\label{bmetric}
ds^2=- \Big(1-\frac{2Mr}{\varrho^2}\Big)dt^2-\frac{4Mr\hat{a}\sin^2\theta}{\varrho^2}
dtd\varphi+\frac{\varrho^2}{\Delta_r}dr^2+\varrho^2d\theta^2
+\frac{A\sin^2\theta}{\varrho^2} d\varphi^2,
\end{eqnarray}
with
\begin{eqnarray}
\varrho^2=r^2+\hat{a}^2\cos^2\theta,\;\;\Delta_\alpha=\frac{r^2-2Mr+\hat{a}^2}{\alpha+1},\;\;A=\bigg(r^2+\hat{a}^2\bigg)^2-(r^2-2Mr+\hat{a}^2)\hat{a}^2
\sin^2\theta~,
\end{eqnarray} where reveals a slight deviation from the standard Kerr.
As a quantity of the underlying background that is expected to appear in the rest of the
paper, should be noted to the horizon angular velocity i.e. $\hat{\Omega}_h= -{\left.\frac{g_{t\phi}}{g_{\phi\phi}}\right|}_{r=r_{eh}}
= \frac{\hat{a}}{r_{eh}^2 +\hat{a}^2}$. Henceforth, to track the role of the mentioned
slight deviation arising from the spontaneously Lorentz symmetry breaking on the black hole superradince, we work
with the metric (\ref{bmetric}) in which the angular momentum, labeled by $\hat{a}$.

\section{Superradiance Scattering of Scalar Wave by Bumblebee Kerr-Black Hole}\label{sec:swb}
The central aim of this paper is probing the trace of the Lorentz-violating parameter $\alpha$
on the superradiance amplification factors which, in essence, address the amount of energy
extracted from a Kerr-like background (\ref{bmetric}) due to a massive scalar field scattering.
We are looking to do this through an analytical treatment.

\subsection{Equation of motion in Bumblebee Kerr background}
Because of our interest in superradiance scattering of a scalar field $\Phi$
with mass $\mu$, thereby, we have to focus on the solving of the Klein-Gordon equation
\begin{equation}\label{eq:KGEq}
\left(\nabla_\alpha \nabla^\alpha +\mu^2\right)\, \Phi(t,r,\theta,\phi)
= 0~.
\end{equation}
The standard Boyer-Lindquist coordinates $(t,r,\theta,\phi)$ let us to a
natural separation of equation \eqref{eq:KGEq} via following ansats,
\begin{equation}\label{eq:an}
\Phi(t,r,\theta,\phi)
= F_{\omega lm}(r)\; S_{\omega lm}(\theta)\; e^{-i\omega t}\; e^{-i m\phi }~,~~~l\geq0,~~~-l\leq m\leq l,~~~\omega>0
\end{equation} include the radial function $F_{\omega lm}(r)$ and oblate spheroidal wave function $ S_{\omega lm}(\theta)$
in which the indices $l,~m$ and $\omega$ respectively represent an integer labelling the angular eigenfunctions,
angular quantum number and the positive frequency of the scattering field as measured by an far away observer.

Putting the ansats \eqref{eq:an} into the partial differential equation \eqref{eq:KGEq},
we deal with two separated ordinary differential equations
\begin{subequations}
\begin{align}
&\frac{d}{d r}\, \left(\Delta_\alpha\, \frac{d F_{\omega lm}(r)}{d r}\right)
+\left(
 \frac{{\left(\left(r^2 +\hat{a}^2\right)\omega -\hat{a}\, m\right)}^2}{\Delta_\alpha}
 -\big(\mu^2 r^2 +l(l+1)+\hat{a}^2\omega^2-2m\hat{a}\omega\big)\right) F_{\omega lm}(r)= 0~, \label{eq:ODE_rad}\\
\begin{split}
&\sin\theta\, \frac{d}{d \theta}\,
 \left(\sin\theta\; \frac{d S_{\omega lm}(\theta)}{d \theta}\right)
 +\bigg(l(l+1)\sin^2\!\theta-\bigg({(\hat{a}\, \omega\, \sin^2\theta -m)}^2
 +\hat{a}^2\, \mu^2\,\sin^2\theta\cos^2\!\theta \bigg)\bigg) S_{\omega lm}(\theta) = 0~. \label{eq:ODE_ang}
\end{split}
\end{align}
\end{subequations}
In direction of our aim, from now on we have to leave the angular equation \eqref{eq:ODE_ang} and concentrate
on the radial equation \eqref{eq:ODE_rad}. Applying a Regge-Wheeler-like coordinate $r_*$ as
\begin{equation}
r_*\equiv\int dr~ \frac{r^2 +\hat{a}^2}{\Delta_{\alpha}}~,~~\big(r_* \rightarrow -\infty~~~\mbox{at event horizon},~~~r_* \rightarrow\infty~~~\mbox{at infinity}\big)
\end{equation} along with introducing a new radial function $\mathcal{F}_{\omega lm}(r_*)= \sqrt{r^2+\hat{a}^2}\, F_{\omega lm}(r)$,
after some calculation one can finally write a Schr\"odinger-like form of deferential equation
\begin{equation} \label{eq:ODE_rad_Tort}
\frac{d^2\mathcal{F}_{\omega lm}(r_*)}{dr_*^2}\,
+U_{\omega lm}(r)\, \mathcal{F}_{\omega lm}(r_*)
= 0~,
\end{equation}
with the following scattering potential
\begin{equation} \label{eq:PotT}
U_{\omega lm}(r)
= {\left(\omega -\frac{m \hat{a}}{r^2 +\hat{a}^2}\right)}^2
-\frac{\Delta_\alpha}{(r^2+\hat{a}^2)^2}\, \bigg(l(l+1)+\hat{a}^2\omega^2-2m\hat{a}\omega+\mu^2 r^2\bigg) \,
+\frac{2\Delta_\alpha(\hat{a}^2-M r)}{(r^2+\hat{a}^2)^3}-\frac{3\hat{a}^2\Delta_\alpha^2}{(r^2+\hat{a}^2)^4}
~.
\end{equation}
Given the fact that the boundary conditions have a central role in the study of scattering
processes, so we have to construct sets of basis modes for the underlying massive scalar field
subject to proper boundary conditions at the event horizon and spatially infinity.
Concerning the asymptotic treatments of the scattering potential in equation~\eqref{eq:PotT},
we come to constant values,
\begin{equation}
\lim_{r\rightarrow r_{eh}} U_{\omega lm}(r)
= {\left(\omega -m\,\tilde{\Omega}_h\right)}^2
\equiv k_{eh}^2~,
\end{equation}
and
\begin{equation}
\lim_{r\to\infty} U_{\omega lm}(r)
= \omega^2 - \lim_{r\to\infty}\, \frac{\mu^2 r^2\, \Delta_\alpha}%
{{\left(r^2 +\hat{a}^2\right)}^2}
= \omega^2 -\hat{\mu}^2\equiv k_\infty^2,~~~\hat{\mu}=\frac{\mu}{\sqrt{\alpha+1}}
\end{equation} at two extremal points horizon and infinity, respectively.
Solving the equation ~\eqref{eq:ODE_rad_Tort} asymptotically, the full radial solution
take the following form
\begin{equation} \label{eq:fullsolution}
\begin{split}
F_{\omega lm}(r)\to
\begin{cases}
\frac{\mathcal{A}_{in}^{eh}\,
  e^{-ik_{eh}\, r_* }}{\sqrt{r_{eh}^2 +\hat{a}^2}}
& \text{for $r \to r_{eh}$}\\
\mathcal{A}_{in}^{\infty}
\frac{e^{-ik_{\infty}\, r_*}}{r}\,
   + \mathcal{A}_{ref}^{\infty}\frac{e^{ik_{\infty}\, r_*}}{r}\,
  & \text{for $r \to \infty$}
\end{cases}
\end{split}
\end{equation}
The indices of the amplitudes $\mathcal{A}$, address the incoming (``in'') or reflected (``ref'') parts
of the scalar wave at the event horizon (``eh'') or at infinity (``$\infty$''). Note that in the above asymptotically
solution at the event horizon there was a reflected wave across the surface $r=r_{eh}$ which usually due to the
presence of a one-sided membrane as horizon and also well-posed Cauchy problem, it is thrown away. In this point,
by facing the Wronskian of the regions near the event horizon $W_{eh}=\big(F_{\omega lm}^{eh}\frac{d F_{\omega lm}^{*~eh}}{dr_*}
-F_{\omega lm}^{*~eh}\frac{d F_{\omega lm}^{eh}}{dr_*}\big)$
and at infinity $W_{\infty}=\big(F_{\omega lm}^{\infty }\frac{d F_{\omega lm}^{*~\infty }}{dr_*}-F_{\omega lm}^{*~\infty }
\frac{d F_{\omega lm}^{\infty }}{dr_*}\big)$
together, we finally have
\begin{eqnarray}\label{flat}
|\mathcal{A}_{ref}^{\infty}|^2=|\mathcal{A}_{in}^{\infty}|^2-\frac{k_{eh}}
{k_\infty}|\mathcal{A}_{in}^{eh}|^2\,,
\end{eqnarray} where is completely free of the details of the scattering potential in Schr\"odinger-like differential equation.
Above equality openly tells us that in case of $\frac{k_{eh}}{k_\infty}<0$ i.e. $\omega<m\hat{\Omega}_{eh}$,
then the scalar wave is superradiantly amplified, $|\mathcal{A}_{ref}^{\infty}|^2>|\mathcal{A}_{in}^{\infty}|^2$.

\subsection{Computation of scalar superradiance amplification factors $Z_{lm}$}
Because the radial equation \eqref{eq:ODE_rad} is not solvable analytically so we have to
resort to some semi-analytically methods. One of the most widely used approximate methods is
so-called asymptotic matching technique which dates back to seminal work of Starobinsky at early 80's
\cite{Starobinsky:1973aij}. So, despite the absence of an exact solution for a singularly perturbed
differential equation as \eqref{eq:ODE_rad}, one still be able to construct an approximation of
the solution via asymptotic expansions in relevant extremal points. Based on this
method actually one find two approximate solutions each one is valid for part of the range of the independent
variable so that finally by their combining together, obtain a trustable single approximate solution.
The underlying technique includes two limiting assumptions: $\hat{a} \omega\ll1$ i.e. $\mathcal{O}(\hat{a}\omega)$
and $M\omega\ll1$ i.e. $\mathcal{O}(M\omega)$ or $\mathcal{O}(\hat{\mu} M)$ \cite{Detweiler:1980uk}.
Former restrict us to low frequency along with slowly rotating regimes for compound system of scalar field
and Kerr like black hole.
Latter forces us to considering the small black hole so that the relevant Compton wavelength of the massive
scalar field is bigger than the size of the black hole.
Technically speaking, by applying the asymptotic matching method in a Kerr background, we are deal with
two zones: region near the  horizon ($r-r_{eh}\ll \omega^{-1}$) called the \emph{``near-region''}, and region
away from the horizon ($r-r_{eh}\gg M$) called the \emph{``far-region''}, outside the event horizon.
Given that matching is possible only when the relevant expansions have a domain of overlap so
the exact solutions derived for the above two asymptotic regions are matched
in an overlapping region where $M\ll r-r_{eh}\ll \omega^{-1}$.

Throughout this subsection, we intend to compute the ``amplification factor'' of a massive scalar wave scattering
off a Bumblebee Kerr black hole through the method discussed above. By this means, we are able to detect the
effect of spontaneously Lorentz symmetry breaking addressed by parameter $\alpha$ on the
amplification factor $Z_{lm}\equiv \frac{|\mathcal{A}_{ref}^{\infty}|^2}{|\mathcal{A}_{in}^{\infty}|^2}-1$,
a dimensionless criterion which represents a black hole superradiance, if $Z_{lm}>0$.
In what follows, by applying the above discussed semi-analytically technique,
we solve the radial equation \eqref{eq:ODE_rad} to deriving $Z_{lm}$.

First, Let's rewrite the radial equation \eqref{eq:ODE_rad} in the form below
\begin{align}\label{eq:ODE_rad2}
&\Delta_\alpha^2\frac{d^2F_{\omega lm}(r)}{d r^2}+\Delta_\alpha\frac{d \Delta_\alpha}{d r}\,. \frac{d F_{\omega lm}(r)}{d r} \nonumber \\
&+\bigg(\big(\left(r^2 +\hat{a}^2\right)\omega -\hat{a}\, m\big)^2 -\Delta_\alpha \big(\mu^2 r^2 +l(l+1)+\hat{a}^2\omega^2-2m\hat{a}\omega\big)\bigg) F_{\omega lm}(r)=0
\end{align}
Our strategy for solving the above differential equation consists of three parts: \textbf{i}) deriving the near-region solution, \textbf{ii}) deriving the far-region solution and \textbf{iii}) matching solutions to having a single solution.
First of all, the equation \eqref{eq:ODE_rad2} after applying the change of variable $x=\frac{r-r_h}{r_h-r_c}$
along with plugging $\triangle_\alpha \frac{d}{dr}=\frac{(r_{eh}-r_{ch})}{\alpha+1}x(x+1) \frac{d}{dx}$ and also using approximation
$\hat{a}\omega\ll1$, reads as
\begin{align}
\label{eq:nr}
&\frac{x^{2}(x+1)^2}{(\alpha+1)^2}\frac{\mathrm{d}^2F_{\omega lm}(x)}{\mathrm{d}x^2}+\frac{x(x+1)(2x+1)}{(\alpha+1)^2}
\frac{\mathrm{d}F_{\omega lm}(x)}{\mathrm{d}x}  \nonumber\\
&+\left(A x^4+B^2-\frac{l(l+1)}{\alpha+1} x(x+1)-\frac{\hat{\mu}^2 A^2}{\omega^2}x^3(x+1)
-\hat{\mu}^2 r_{eh}^2 x(x+1)-\frac{2\hat{\mu}^2 r_{eh}^2 A}{\omega} x^2(x+1)\right) F_{\omega lm}(x)=0\,,
\end{align} where $A=(r_{eh}-r_{ch})\omega$ and $B=\frac{(\omega-m\hat{\Omega})}{r_{eh}-r_{ch}}r_{eh}^2$.
By focusing on part \textbf{i}, the equation \eqref{eq:nr} due to two applicable approximations in regions close to
the horizon i.e. $Ax\ll$ and $\hat{\mu}^2r_{eh}^2\ll1$, reduces to
\begin{align}\label{eq:nrr}
x^{2}(x+1)^2\frac{\mathrm{d}^2F_{\omega lm}(r)}{\mathrm{d}x^2}+x(x+1)(2x+1)\frac{\mathrm{d}F_{\omega lm}(r)}{\mathrm{d}x}
+\left((\alpha+1)^2B^2-l(l+1)(\alpha+1) x(x+1)\right)F_{\omega lm}(r)=0\,.
\end{align} Note that the second approximation ($\hat{\mu}^2r_{eh}^2\ll1$) comes from Compton wavelength approximation as one of central
assumptions in asymptotic matching technique. The general solution of equation \eqref{eq:nrr} satisfying the ingoing boundary
condition, written in terms of Legendre function of the first kind $P^{\nu}_{\lambda}(y)$
\begin{eqnarray}\label{general}
F_{\omega lm}(x)=c~P^{2 i (\alpha+1)B}_{\frac{\sqrt{1+4(\alpha+1)l(l+1)}-1}{2}}(1+2x)~,
\end{eqnarray} where using following relation
\begin{eqnarray}\label{g}
P^{\nu}_{\lambda}(y)=\frac{1}{\Gamma(1-\nu)}\big(\frac{1+y}{1-y}\big)^{\nu/2}\, _2F_1(-\lambda,\lambda+1;1-\nu;\frac{1-y}{2})~,
\end{eqnarray} it finally re-express in terms of ordinary hypergeometric functions $_2F_1(a,b;c;y)$
\begin{align}
F_{\omega lm}(x)=c~\big(\frac{x}{x+1} \big)^{-i(\alpha+1)~B}~_2F_1\left(\frac{1-\sqrt{1+4(\alpha+1)l(l+1)}}{2},\frac{1+\sqrt{1+4(\alpha+1)l(l+1)}}{2};1-2i(\alpha+1)~B;-x\right).
\end{align}
Matching of solutions requires us to looking for the manner of above solution at large $x$ i.e.
\footnote{ To obtain equation \eqref{eq:solnear}, used from asymptotic behaviour of the hypergeometric function
$\lim\limits_{x \rightarrow \infty}~ _2F_1(a,b;c;-y) = \dfrac{\Gamma(b-a)\Gamma(c)}{\Gamma(c-a)\Gamma(b)}y^{-a}+\dfrac{\Gamma(a-b)\Gamma(c)}{\Gamma(c-b)\Gamma(a)}y^{-b}$}
\begin{align} \label{eq:solnear}
F_{\mathrm{near- large~x}}\sim c~\left(\frac{\Gamma\big(\sqrt{1+4(\alpha+1)l(l+1)}\big)~\Gamma\big(1-2i(\alpha+1)B\big)}{\Gamma\bigg(\frac{1+\sqrt{1+4(\alpha+1)l(l+1)}}{2}
-2i(\alpha+1)B\bigg)
~\Gamma\bigg(\frac{1+\sqrt{1+4(\alpha+1)l(l+1)}}{2}\bigg)}
~x^{\frac{\sqrt{1+4(\alpha+1)l(l+1)}-1}{2}}+ \right.\nonumber\\\left.
\frac{\Gamma\big(-\sqrt{1+4(\alpha+1)l(l+1)}\big)~\Gamma\big(1-2i(\alpha+1)~B\big)}{\Gamma\bigg(\frac{1-\sqrt{1+4(\alpha+1)l(l+1)}}{2}\bigg)
~\Gamma\bigg(\frac{1-\sqrt{1+4(\alpha+1)l(l+1)}}{2}-2i(\alpha+1)B\bigg)}~x^{-\frac{\sqrt{1+4(\alpha+1)l(l+1)}+1}{2}} \right)\,.
\end{align}
\\
By going to part \textbf{ii} of strategy in order to deriv the far-region solution, equation \eqref{eq:ODE_rad2}
reads as
\begin{eqnarray}\label{eq:far}
\frac{\mathrm{d}^2F_{\omega lm}(x)}{\mathrm{d}x^2}+\frac{2}{x}\frac{\mathrm{d}F_{\omega lm}(x)}{\mathrm{d}x}+
\left(k^2-\frac{l(l+1)(\alpha+1)}{x^2}\right)F_{\omega lm}(x)=0\,,
\end{eqnarray} where $k \equiv \frac{A}{\omega}\sqrt{\omega^2-\hat{\mu}^2}$.
Note that here we used from approximations $x+1\approx x$ and $\hat{\mu}^2r_{eh}^2\ll1$.
For the equation \eqref{eq:far}, we have following general solution
\begin{align}\label{sol}
F_{\omega lm,~\mathrm{far}} = e^{-i k x} \left(d_1~x^{\frac{\sqrt{1+4(\alpha+1)l(l+1)}-1}{2}}~U\big(\frac{1+\sqrt{1+4(\alpha+1)l(l+1)}}{2},1+\sqrt{1+4(\alpha+1)l(l+1)},2ikx\big)+
\right.\nonumber\\\left.
d_2~x^{-\frac{\sqrt{1+4(\alpha+1)l(l+1)}-1}{2}}~U\big(\frac{1-\sqrt{1+4(\alpha+1)l(l+1)}}{2},1-\sqrt{1+4(\alpha+1)l(l+1)},2ik x\big)\right)\,,
\end{align} where $U(a,b,y)$ refers to the first Kummer function.
Matching of solutions requires us to looking for the manner of above solution at small $x$ i.e. \footnote{Here used from
the Taylor expansion $\lim_{y\rightarrow0}U(a,b,y)\approx\frac{\Gamma(1-b)}{\Gamma(1+a-b)}+....$ }
\begin{eqnarray}\label{eq:solfar}
F_{\omega lm,~\mathrm{far-small \,\, x}}\sim  d_1~x^{\frac{\sqrt{1+4(\alpha+1)l(l+1)}-1}{2}}+
d_2~x^{-\frac{1+\sqrt{1+4(\alpha+1)l(l+1)}}{2}}\,.
\end{eqnarray}
Now we are in suitable point to apply part \textbf{iii} of strategy i.e. matching above derived two asymptotic solutions
to calculate the scalar wave fluxes at infinity. In this way, we will have the relevant expression for the amplification factor.
At first step, by facing equations \eqref{eq:solnear} and \eqref{eq:solfar} together we acquire
\begin{eqnarray} \label{eq:d12}
d_1&=& c\frac{\Gamma(\sqrt{1+4(\alpha+1)l(l+1)})\Gamma(1-2i(\alpha+1)B)}{\Gamma(\frac{1+\sqrt{1+4(\alpha+1)l(l+1)}}{2})-2i(\alpha+1)B)
\Gamma(\frac{1+\sqrt{1+4(\alpha+1)l(l+1)}}{2})}\,,\\ \nonumber
d_2&=&c\frac{\Gamma(-\sqrt{1+4(\alpha+1)l(l+1)})\Gamma(1-2i(\alpha+1)B)}{\Gamma(\frac{1-\sqrt{1+4(\alpha+1)l(l+1)}}{2})-2i(\alpha+1)B)
\Gamma\big(\frac{1-\sqrt{1+4(\alpha+1)l(l+1)}}{2}\big)}\,.
\end{eqnarray}
Now we need to connect the coefficients $d_1$ and $d_2$ with coefficients $\mathcal{A}_{in}^{\infty}$ and
$\mathcal{A}_{ref}^{\infty}$ in the radial solution \eqref{eq:fullsolution}.
To do this, we expand the far region solution \eqref{sol} around infinity as
\begin{align}\label{eq:Asysoo}
d_1 \frac{\Gamma(1+\sqrt{1+4(\alpha+1)l(l+1)})}{\Gamma(\frac{1+\sqrt{1+4(\alpha+1)l(l+1)}}{2})}k^{-\frac{1+\sqrt{1+4(\alpha+1)l(l+1)}}{2}}
\bigg((-2i)^{-\frac{1+\sqrt{1+4(\alpha+1)l(l+1)}}{2}}\frac{e^{-ik x}}{x}+
(2i)^{-\frac{1+\sqrt{1+4(\alpha+1)l(l+1)}}{2}}\frac{e^{ik x}}{x}\bigg)+\\ \nonumber
d_2 \frac{\Gamma(1-\sqrt{1+4(\alpha+1)l(l+1)})}{\frac{1-\sqrt{1+4(\alpha+1)l(l+1)}}{2}}k^{\frac{\sqrt{1+4(\alpha+1)l(l+1)}-1}{2}}
\bigg((-2i)^{\frac{\sqrt{1+4(\alpha+1)l(l+1)}-1}{2}}\frac{e^{-ik x}}{x}+
(2i)^{\frac{\sqrt{1+4(\alpha+1)l(l+1)}-1}{2}}\frac{e^{ik x}}{x}\bigg)\, .
\end{align}
Using the approximations $\frac{1}{x}\sim\frac{A}{\omega}.\frac{1}{r},~~
e^{\pm i k x}\sim e^{\pm i\sqrt{\omega^2-\tilde{\mu}^2} r}$ and then matching the solution
\eqref{eq:Asysoo} with the radial solution \eqref{eq:fullsolution}
\begin{align}\label{eq:Asysooo}
F_\infty(r)\sim \mathcal{A}_{in}^{\infty}~\frac{e^{-i\sqrt{\omega^2-\tilde{\mu}^2} r^*}}{r}+\mathcal{A}_{ref}^{\infty}~
\frac{e^{i\sqrt{\omega^2-\tilde{\mu}^2} r^*}}{r}, \qquad  \mbox{for}~~~r \rightarrow \infty ,
\end{align} we have
\begin{eqnarray}\label{eq:A1}
\mathcal{A}_{in}^{\infty}=\frac{A}{\omega}\left(d_1 (-2i)^{-\frac{1+\sqrt{1+4(\alpha+1)l(l+1)}}{2}}~
\frac{\Gamma(1+\sqrt{1+4(\alpha+1)l(l+1)})}{\Gamma(\frac{1+\sqrt{1+4(\alpha+1)l(l+1)}}{2})}~k^{-\frac{1+\sqrt{1+4(\alpha+1)l(l+1)}}{2}}
+\right.\nonumber\\\left.
d_2 (-2i)^{\frac{\sqrt{1+4(\alpha+1)l(l+1)}-1}{2}}~
\frac{\Gamma(1-\sqrt{1+4(\alpha+1)l(l+1)})}{\Gamma(\frac{1-\sqrt{1+4(\alpha+1)l(l+1)}}{2})}~k^{\frac{\sqrt{1+4(\alpha+1)l(l+1)}-1}{2}}\right)\,,
\end{eqnarray}
and
\begin{eqnarray}\label{eq:A2}
\mathcal{A}_{ref}^{\infty}=\frac{A}{\omega}\left(d_1 (2i)^{-\frac{1+\sqrt{1+4(\alpha+1)l(l+1)}}{2}}~
\frac{\Gamma(1+\sqrt{1+4(\alpha+1)l(l+1)})}{\Gamma(\frac{1+\sqrt{1+4(\alpha+1)l(l+1)}}{2})}~k^{-\frac{1+\sqrt{1+4(\alpha+1)l(l+1)}}{2}}
+\right.\nonumber\\\left.
d_2 (2i)^{\frac{\sqrt{1+4(\alpha+1)l(l+1)}-1}{2}}
\frac{\Gamma(1-\sqrt{1+4(\alpha+1)l(l+1)})}{\Gamma(\frac{1-\sqrt{1+4(\alpha+1)l(l+1)}}{2})}~k^{\frac{\sqrt{1+4(\alpha+1)l(l+1)}-1}{2}}\right)\,.
\end{eqnarray}
As the last step, by adding the expressions obtained in \eqref{eq:d12} for $d_1$ and $d_2$, the final form of
$\mathcal{A}_{ref}^{\infty}$ and $\mathcal{A}_{in}^{\infty}$, written as
\begin{eqnarray}\label{eq:A3}
&&\mathcal{A}_{in}^{\infty}=\frac{c~ (-2i)^{-\frac{1+\sqrt{1+4(\alpha+1)l(l+1)}}{2}}}{\sqrt{\omega^2-\tilde{\mu}^2}}.\frac{
\Gamma(\sqrt{1+4(\alpha+1)l(l+1)})~\Gamma(1+\sqrt{1+4(\alpha+1)l(l+1)})}
{\Gamma\bigg(\frac{1+\sqrt{1+4(\alpha+1)l(l+1)}}{2}-2i(\alpha+1) B\bigg)~\bigg(\Gamma(\frac{1+\sqrt{1+4(\alpha+1)l(l+1)}}{2})\bigg)^2}\times \\ \nonumber
&&~\Gamma(1-2i(\alpha+1)~ B)~k^{\frac{1-\sqrt{1+4(\alpha+1)l(l+1)}}{2}}+ \frac{c~ (-2 i)^{\frac{\sqrt{1+4(\alpha+1)l(l+1)}-1}{2}}}{\sqrt{\omega^2-\hat{\mu}^2}}
\times\\ \nonumber
&&\frac{\Gamma(1-\sqrt{1+4(\alpha+1)l(l+1)})
~\Gamma(-\sqrt{1+4(\alpha+1)l(l+1)})}
{\bigg(\Gamma(\frac{1-\sqrt{1+4(\alpha+1)l(l+1)}}{2})\bigg)^2 ~\Gamma\big(\frac{1-\sqrt{1+4(\alpha+1)l(l+1)}}{2}-2i(\alpha+1)~ B\big)}
~\Gamma(1-2i(\alpha+1)~ B)~k^{\frac{1+\sqrt{1+4(\alpha+1)l(l+1)}}{2}}\,,
\end{eqnarray}
and
\begin{eqnarray}\label{eq:A4}
&&\mathcal{A}_{ref}^{\infty}=\frac{c~(2i)^{-\frac{1+\sqrt{1+4(\alpha+1)l(l+1)}}{2}}}{\sqrt{\omega^2-\hat{\mu}^2}}.\frac{
\Gamma(\sqrt{1+4(\alpha+1)l(l+1)})~\Gamma(1+\sqrt{1+4(\alpha+1)l(l+1)})}
{\Gamma\bigg(\frac{1+\sqrt{1+4(\alpha+1)l(l+1)}}{2}-2i(\alpha+1)~B\bigg)~\bigg(\Gamma(\frac{1+\sqrt{1+4(\alpha+1)l(l+1)}}{2})\bigg)^2}\times
\\ \nonumber
&&\Gamma(1-2i(\alpha+1)~B)~k^{\frac{1-\sqrt{1+4(\alpha+1)l(l+1)}}{2}}+\frac{c~(2i)^{\frac{\sqrt{1+4(\alpha+1)l(l+1)}-1}{2}}}{\sqrt{\omega^2-\hat{\mu}^2}}
\times\\ \nonumber
&&\frac{\Gamma(1-\sqrt{1+4(\alpha+1)l(l+1)})
~\Gamma(-\sqrt{1+4(\alpha+1)l(l+1)})}
{\bigg(\Gamma(\frac{1-\sqrt{1+4(\alpha+1)l(l+1)}}{2})\bigg)^2 ~\Gamma\bigg(\frac{1-\sqrt{1+4(\alpha+1)l(l+1)}}{2}-2i(\alpha+1)~ B\bigg)}
\Gamma(1-2i(\alpha+1)~ B)~k^{\frac{1+\sqrt{1+4(\alpha+1)l(l+1)}}{2}}\,.
\end{eqnarray} respectively. Now one able to the calculation of the amplification factor
\begin{eqnarray}\label{amp}
Z_{lm}\equiv \frac{|\mathcal{A}_{ref}^{\infty}|^2}{|\mathcal{A}_{in}^{\infty}|^2}-1\,,
\end{eqnarray} via equations \eqref{eq:A3} and \eqref{eq:A4}.

\subsection{Results}
\begin{figure}[!ht]
	\begin{center}
			\includegraphics[scale=0.3]{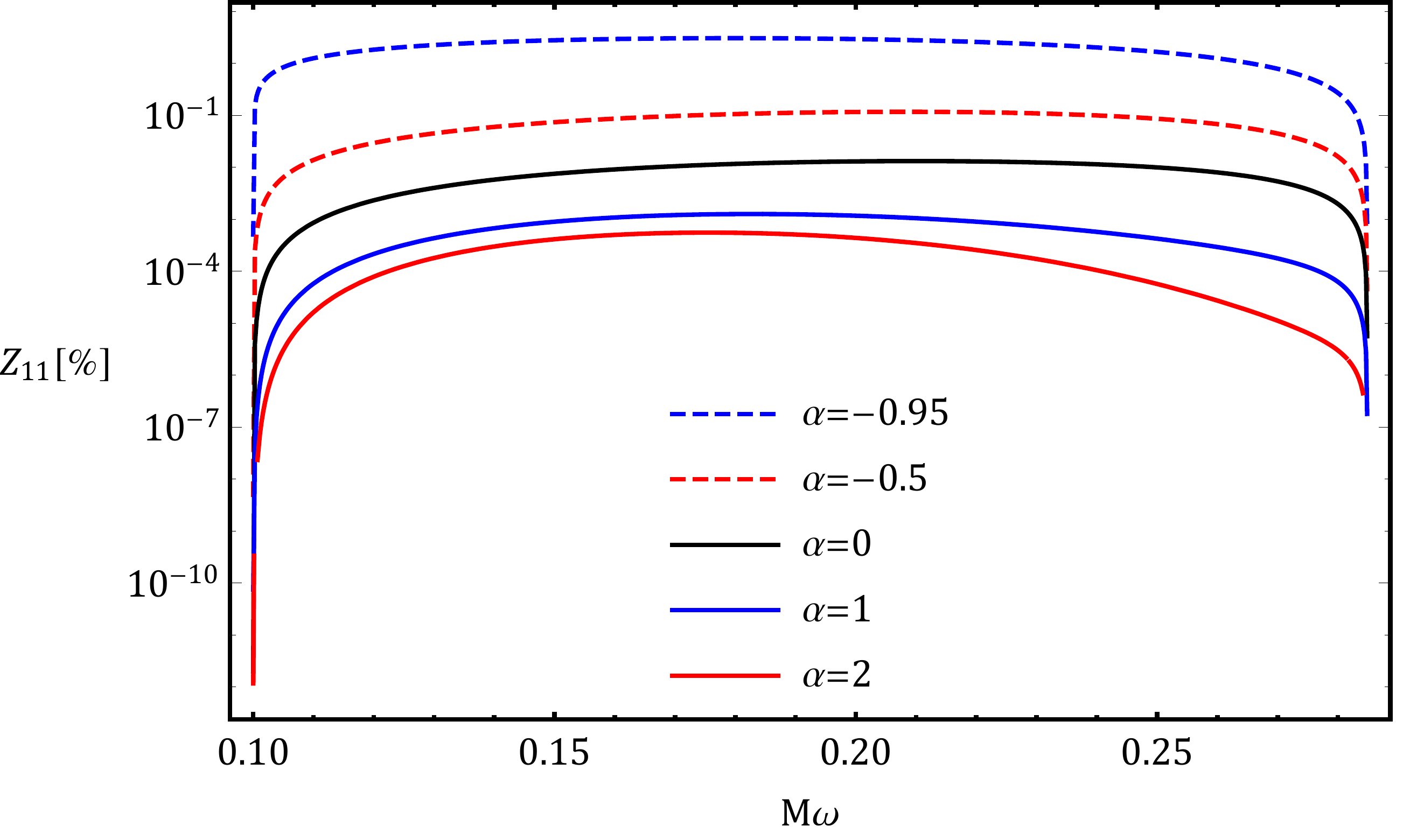}
			\includegraphics[scale=0.3]{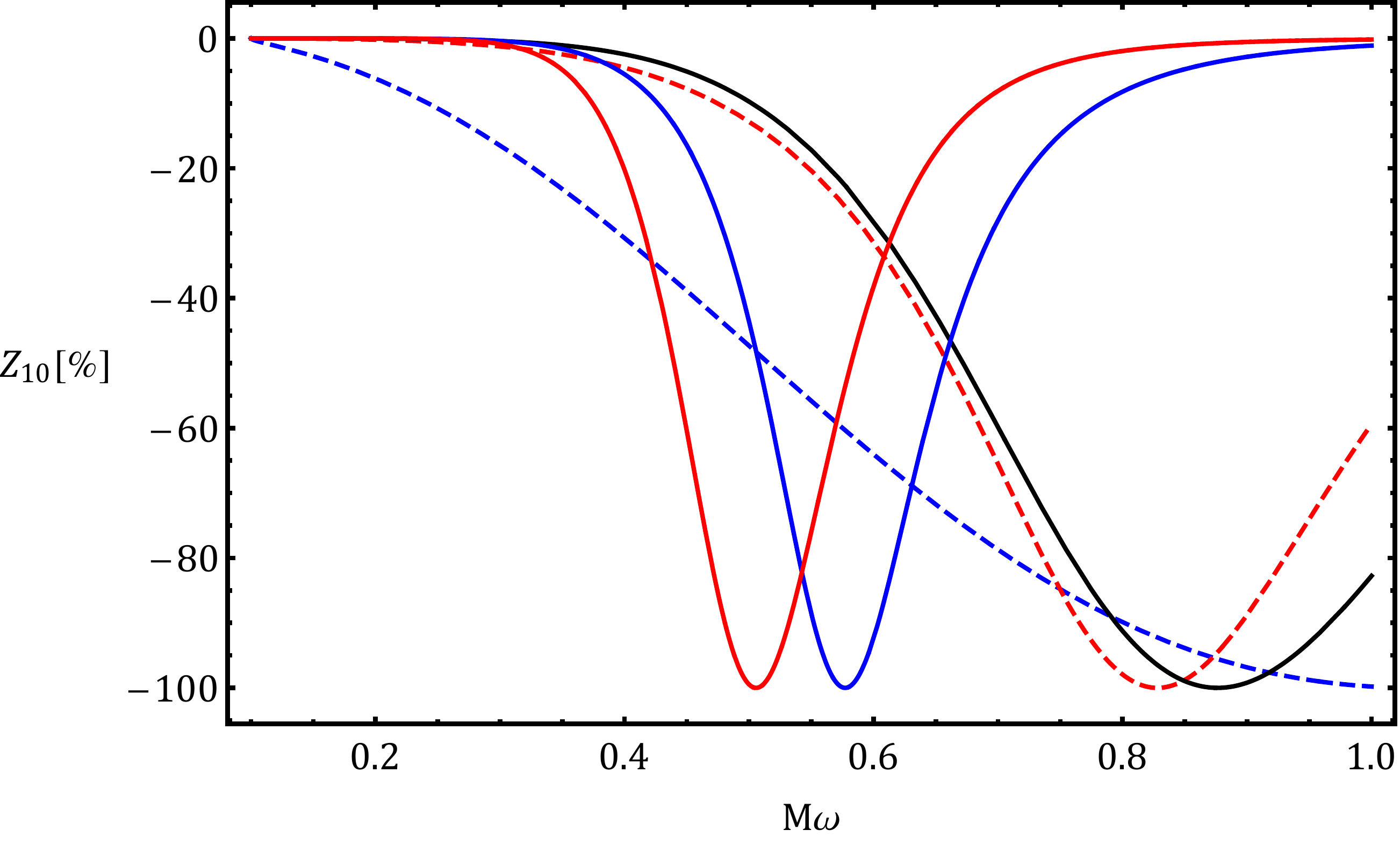}
            \includegraphics[scale=0.3]{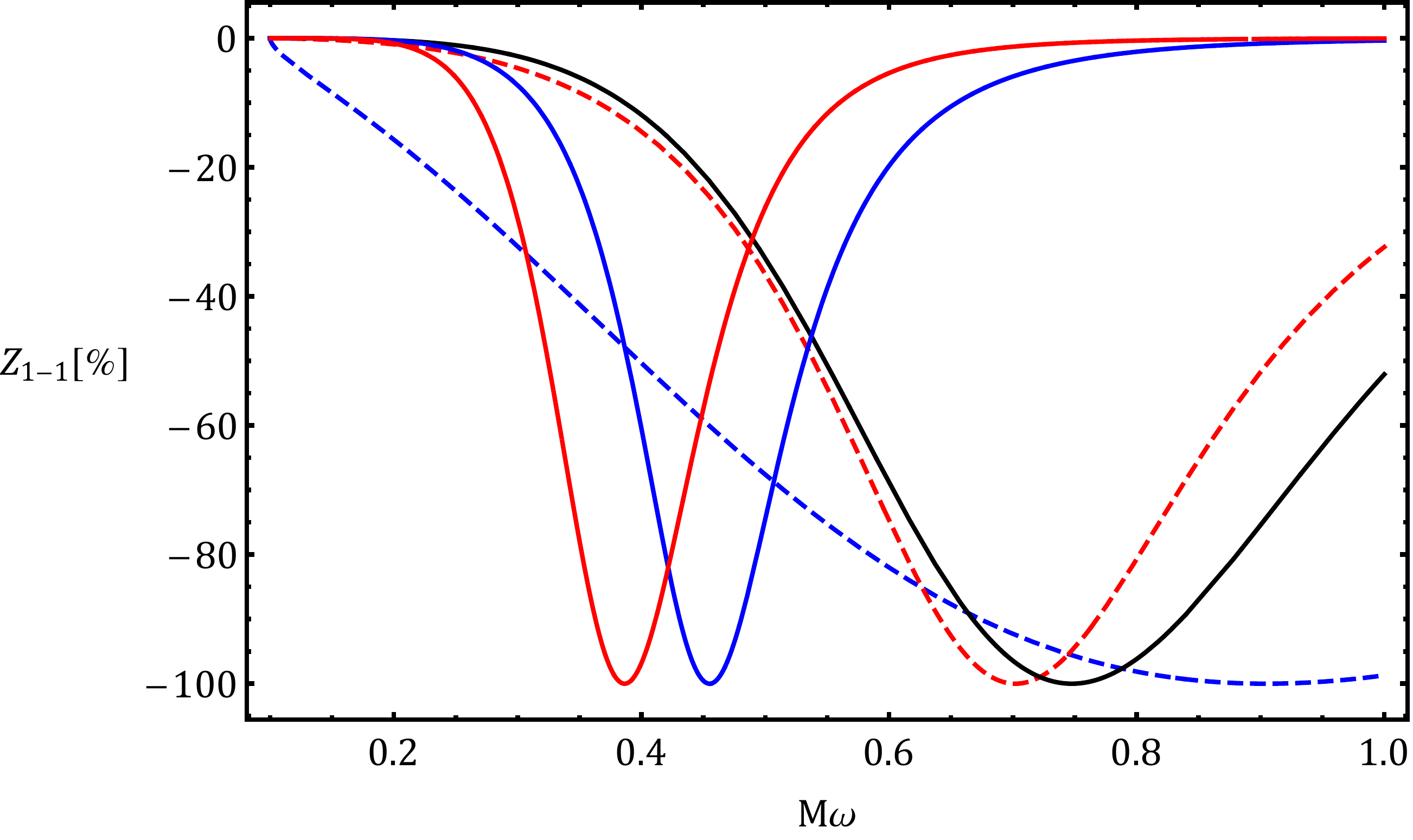}
			 \caption{Analytical results obtained from (\ref{amp}) for the
			 	amplification factor of the massive scalar wave with the multipoles $(l,m)=(1,1),~(1,0),~(1,-1)$ and mass $\hat{\mu}=0.1$
			 	scattered off the bumblebee Kerr like background
with $\hat{a}/M=0.9$ and different values of Lorentz violating parameter $\alpha$.}
		\label{Z1}
	\end{center}
     \end{figure}

\begin{figure}[!ht]
	\begin{center}
			\includegraphics[scale=0.3]{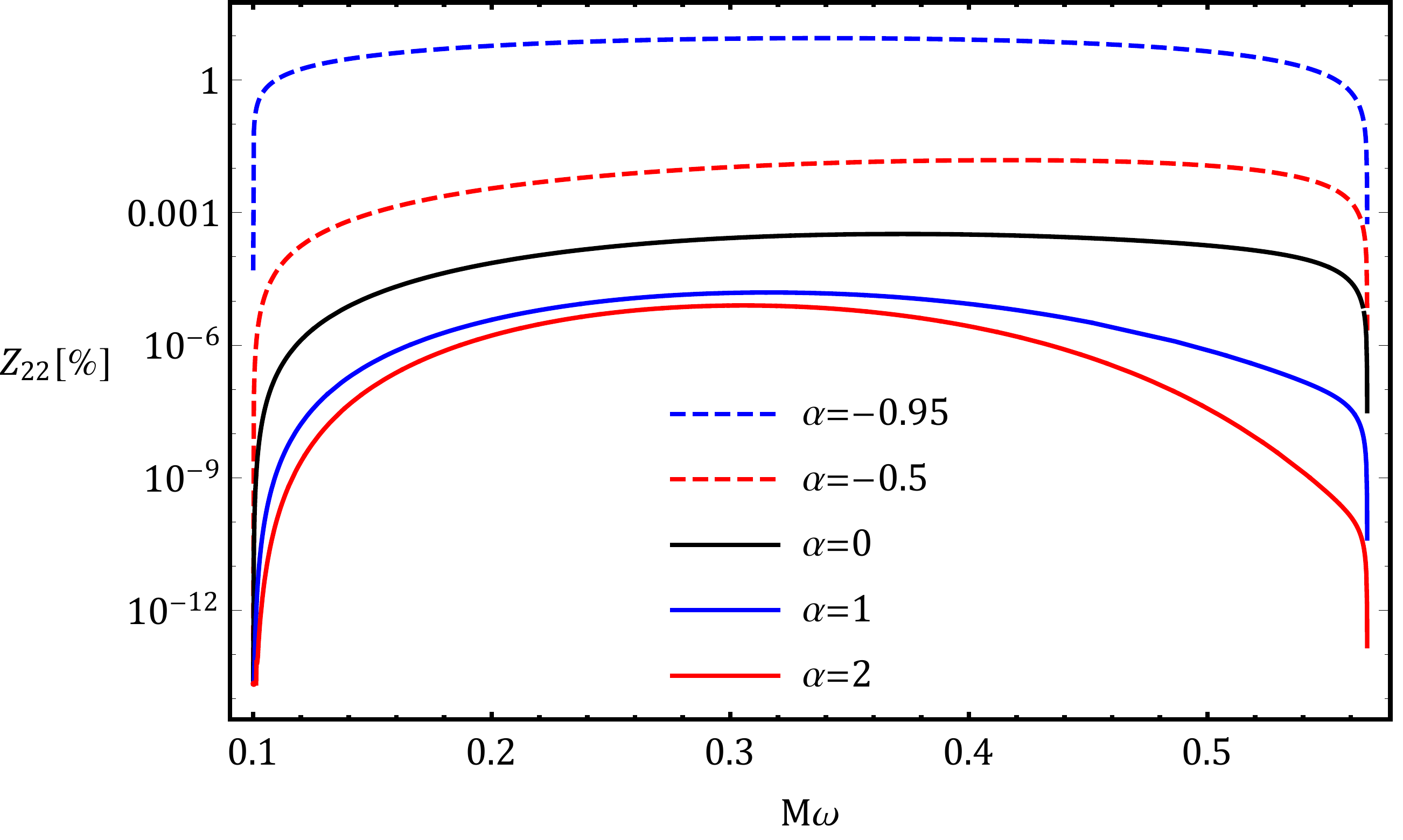}
			\includegraphics[scale=0.3]{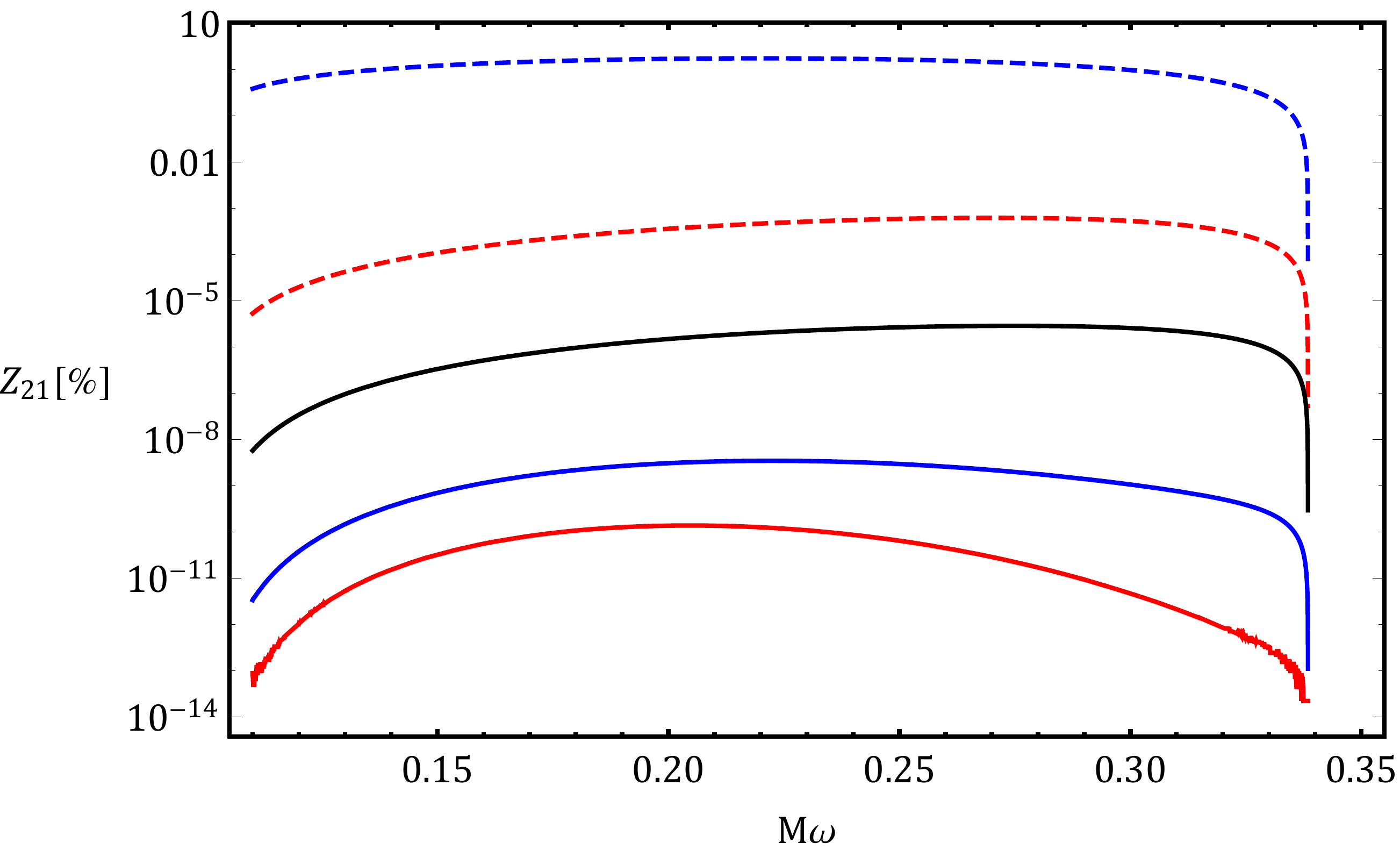}
            \includegraphics[scale=0.3]{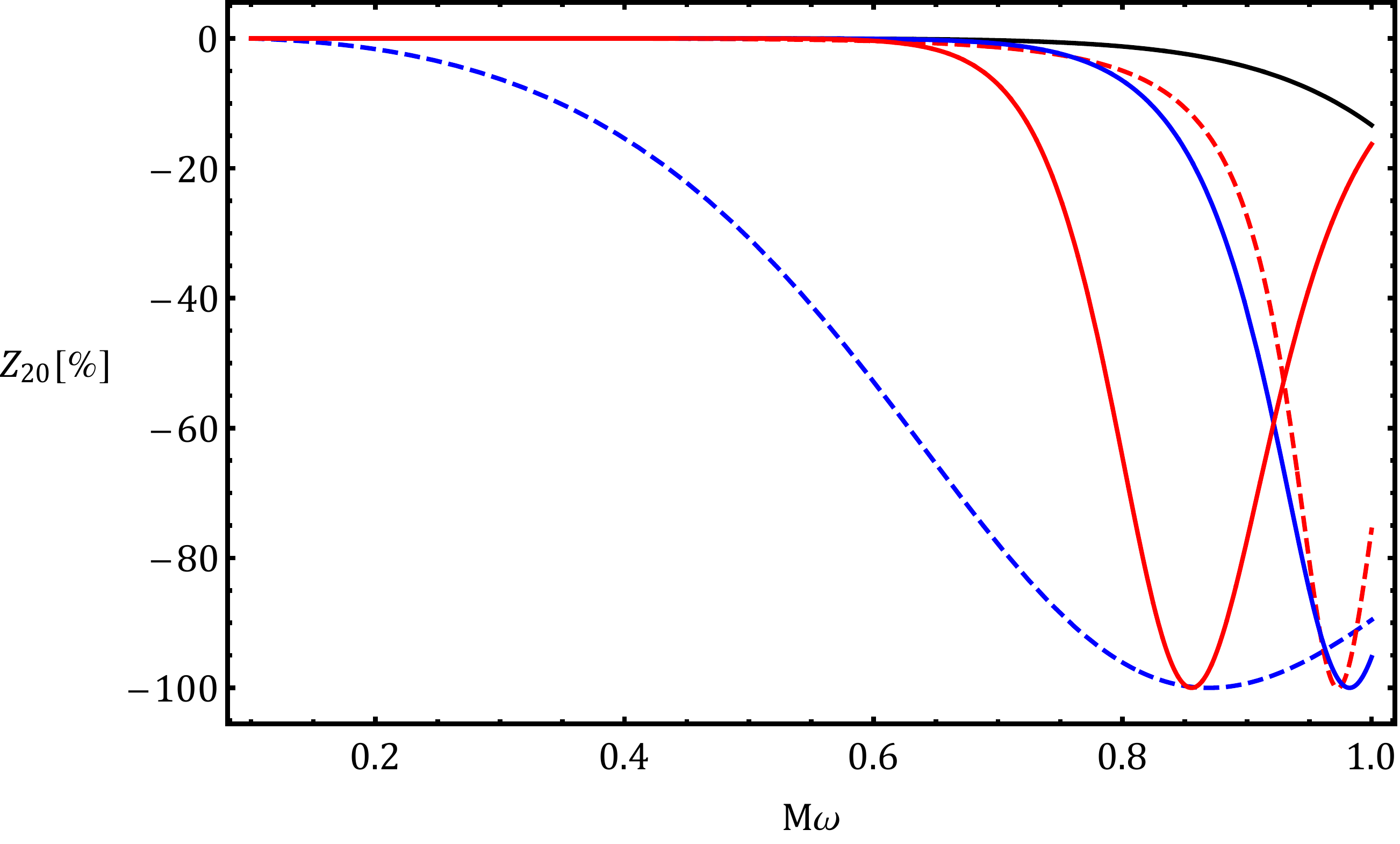}
            \includegraphics[scale=0.3]{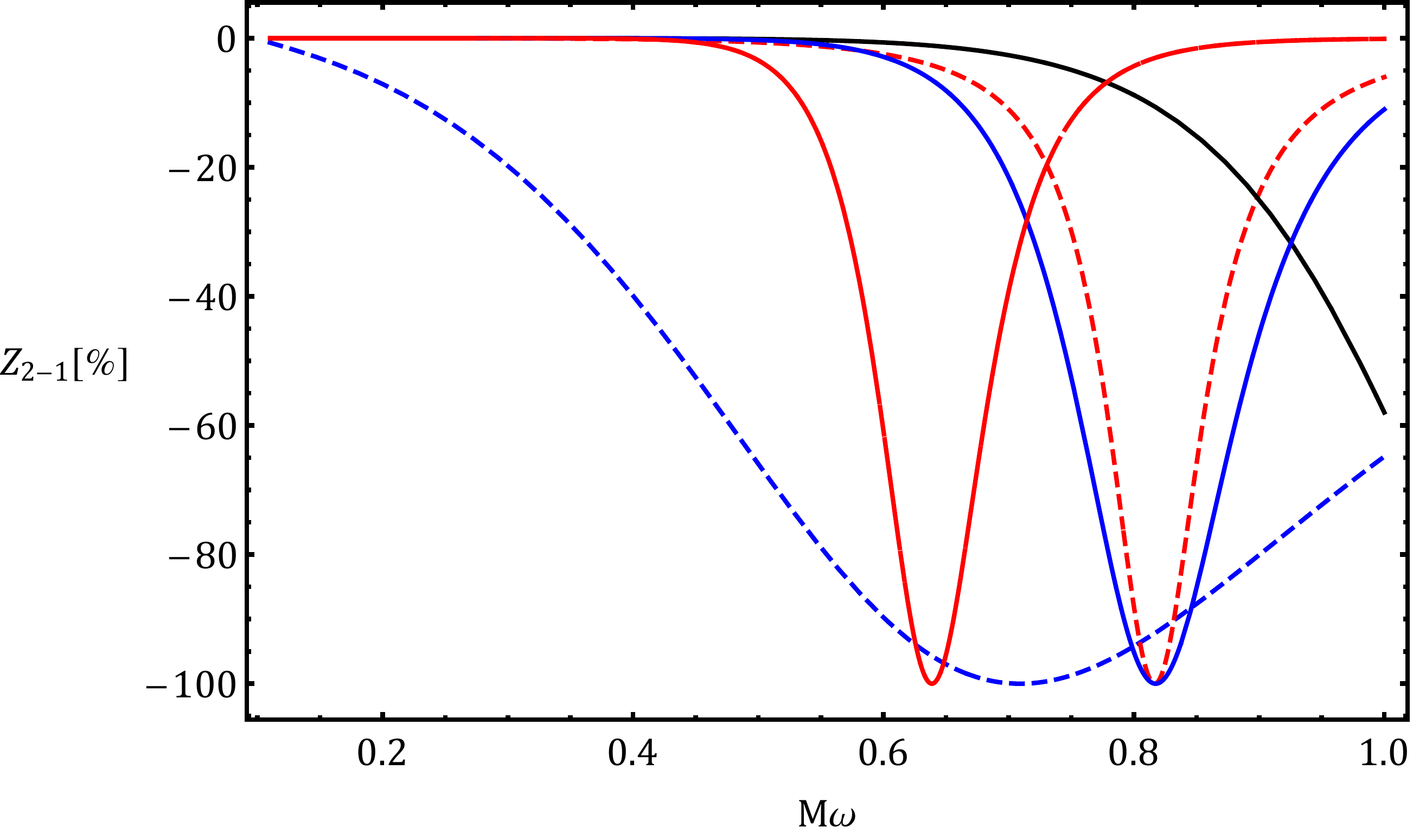}
			\includegraphics[scale=0.3]{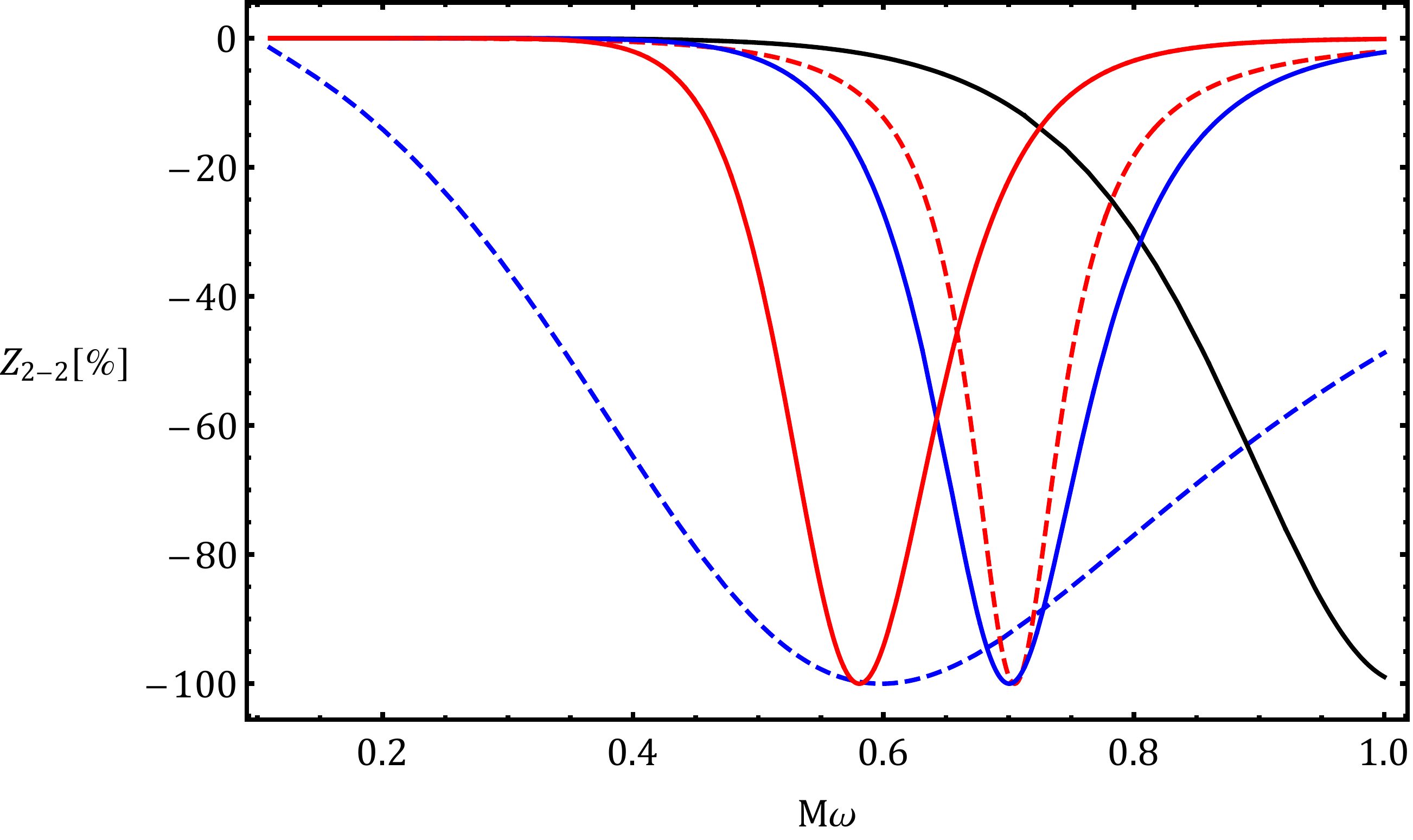}
             \caption{Analytical results obtained from (\ref{amp}) for the
             amplification factor of the massive scalar wave with the multipoles  $(l,m)=(2,2),~(2,1),~(2,0),~(2,-1),~(2,-2)$ and mass $\hat{\mu}=0.1$
             scattered off the bumblebee Kerr like background
             with $\hat{a}/M=0.9$ and different values of Lorentz violating parameter $\alpha$.}
		\label{Z2}
	\end{center}
     \end{figure}

First, we release the results derived above for the leading multipoles ($l=1,2$) of the massive scalar wave scattered off a Kerr black hole with the ratio angular momentum to mass $\hat{a}/M=0.9$, see  Figs. (\ref{Z1}) and (\ref{Z2}). Indeed, if $Z>0$ it addresses a gain factor assigned to scalar wave mode means happening the superradiance  phenomena in the underlying background while $Z<0$ denote a loss factor means the non-occurrence of superradiance.

 \begin{figure}[!ht]
	\begin{center}
			\includegraphics[scale=0.3]{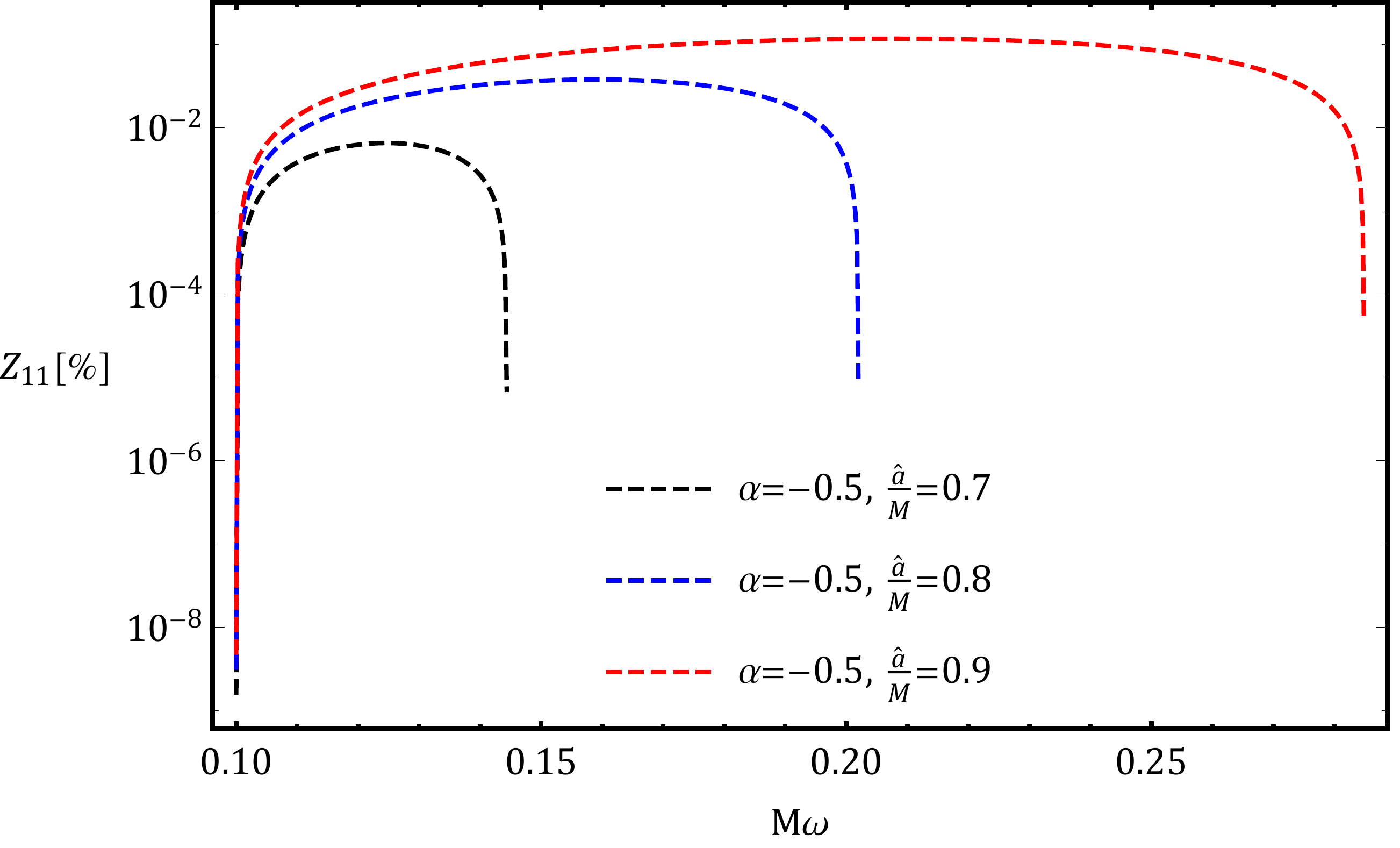}
			\includegraphics[scale=0.3]{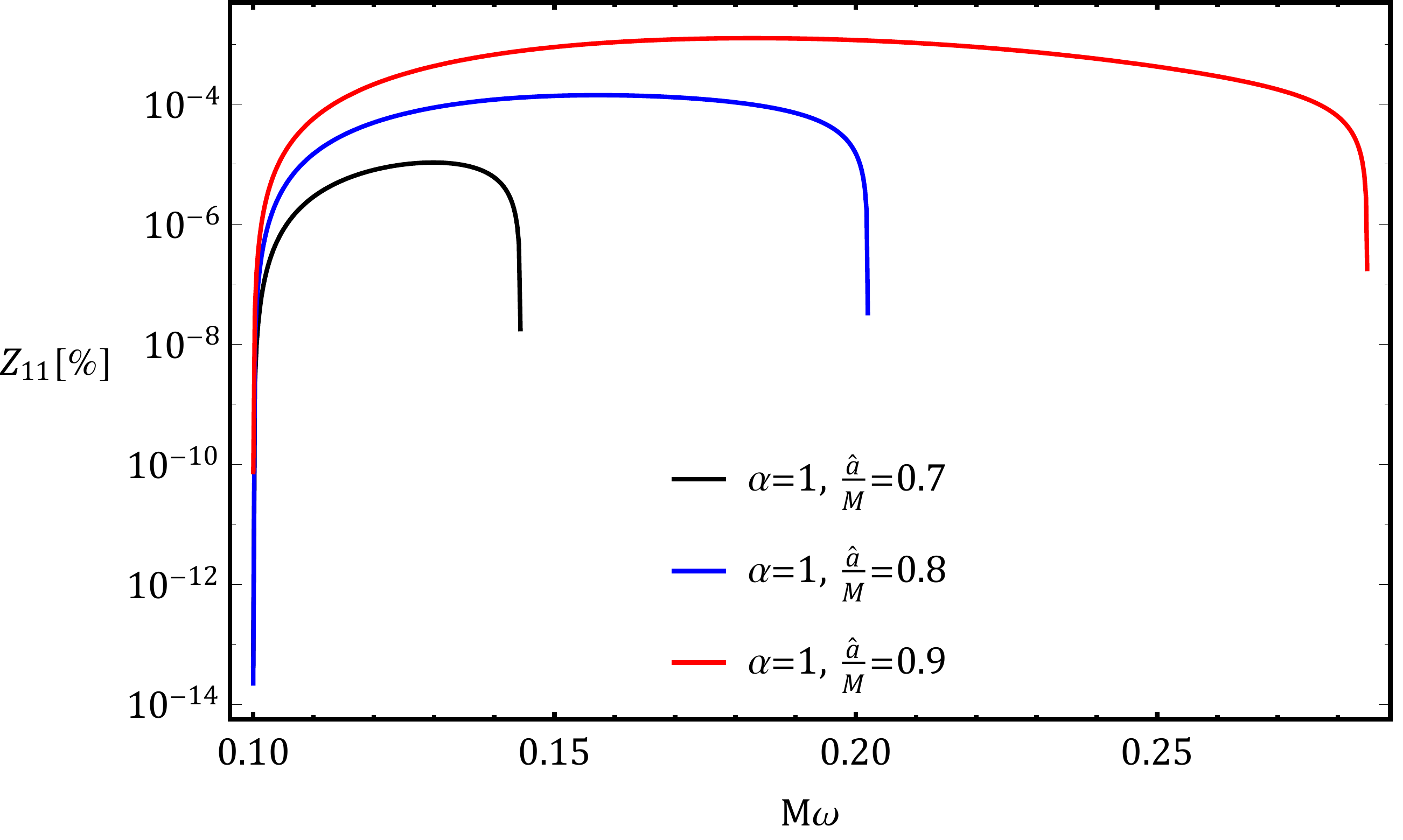}
            \includegraphics[scale=0.3]{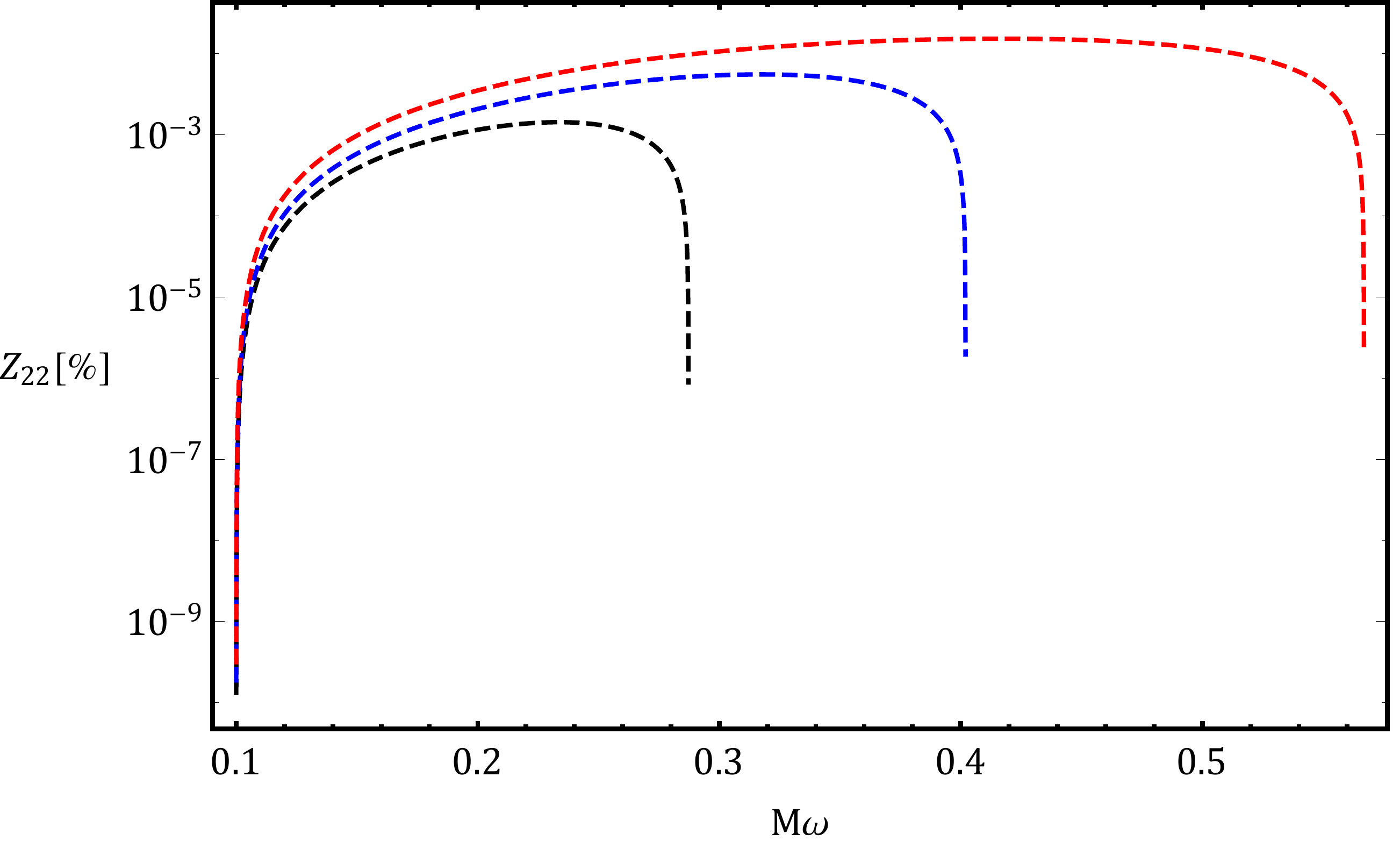}
            \includegraphics[scale=0.3]{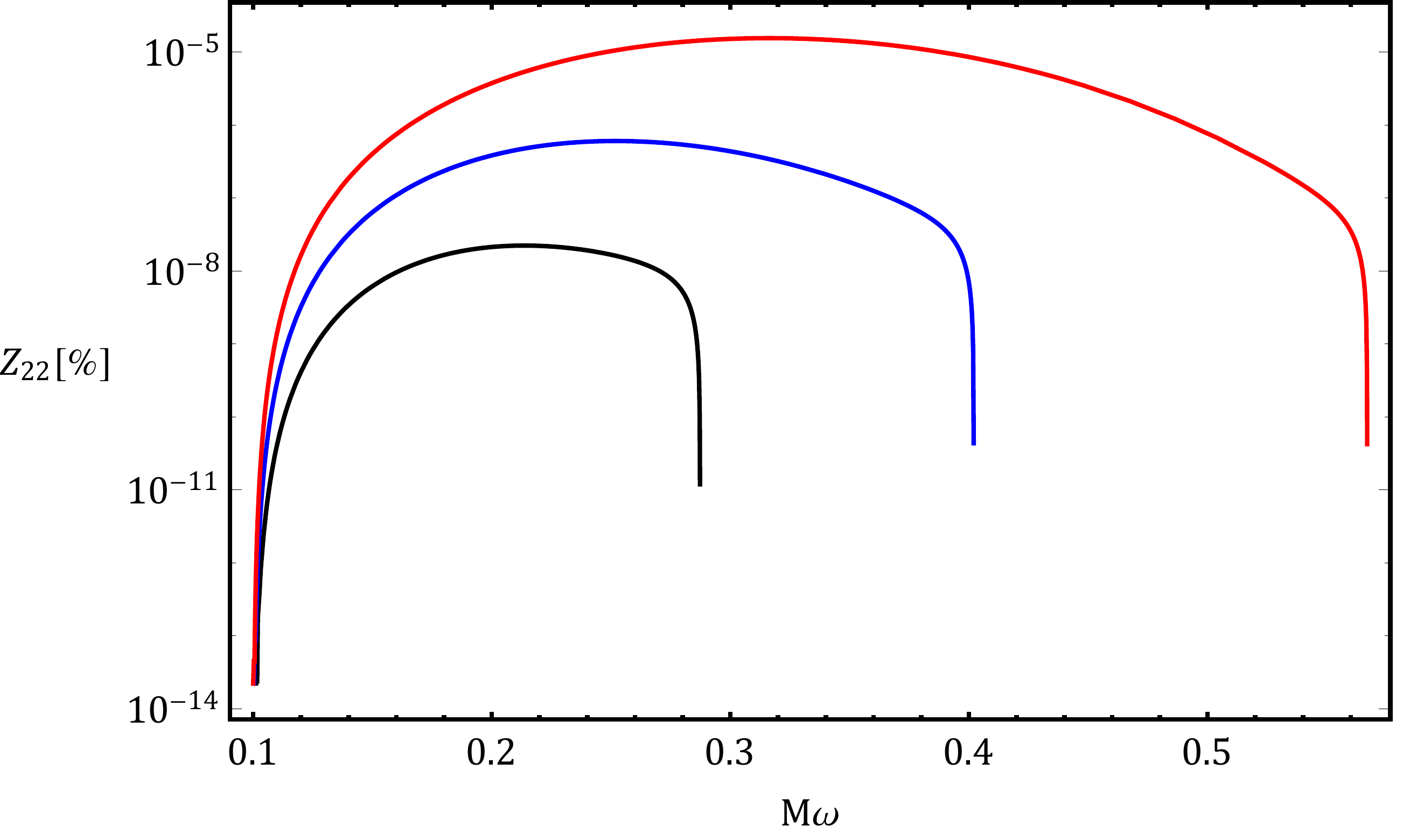}
			 \caption{Amplification factors $Z_{11}$ and $Z_{22}$ for two optional values of $\alpha$: $\alpha=-0.5$ (left panels in top and bottom rows) and $\alpha=1$ (right panels in top and bottom rows) with different values of $\hat{a}/M=0.7,~0.8,~0.9$.}
		\label{Z3}
	\end{center}
     \end{figure}

Concerning these two figures, we can divide our analysis to two cases non-positive $m\leq0$ and positive $m>0$.
In similar to the standard Kerr black hole, for the case $m\leq0$ the massive scalar field includes the non-superradiant
modes since we deal with $Z\leq0$ within the relevant frequency range. However, in the presence of Lorentz violating parameter $\alpha$
we see more wave energy loss relative to the standard case within the frequency regions $M\omega\ll 1$ (as our study validity range).
Actually, the presence of $\alpha $
in the background seems to accelerate the process of energy loss in the non-superradiant modes. However, by going to case $m>0$,
superradiance phenomena will occurs with two distinguishable feedbacks from the Lorentz violating parameter $\alpha$.
The plots related to case $m>0$ in the Figs. (\ref{Z1}) and (\ref{Z2}), clear-cut tell us that
$\alpha<0$ results in enhanced superradiance while $\alpha>0$, weakens it. This may seems more interesting if we confront it
with the result released in \cite{Ding:2019mal} about relation between the Lorentz violating parameter $\alpha$ and the size
of the shadow. There, the results indicate that the shadow's size of the Kerr-like black hole at hand, increases, and decreases compared to the standard case respectively for $\alpha<0$ and $\alpha>0$.
Due to this comparison, this idea is prone to emerge that the observation of a deviation in the size of the shadow in the direction of the increase may be indirect evidence of happening the superradiance phenomenon.

One may want to search for the relation between the black hole's spin $\hat{a}$ and Lorentz violating parameter $\alpha$ with
the superradiance amplification factor. In this direction, by choosing two optional values for negative and positive Lorentz
violating parameter $\alpha$, we release the superradiance amplification factors $Z_{11}$ and $Z_{22}$ in Fig. (\ref{Z3}), for
some values of black hole's spin $\hat{a}$. Generally speaking, for both cases $\alpha<0$ and $\alpha>0$ we observe that
the superradiance frequency range as well as the their amplification factors become wider and stronger, as the black hole's
spin gets bigger. It could contain the message that as the Kerr black hole gets closer to the extremal case $\hat{a}=M$, the effects
arising from Lorentz symmetry breaking on the superradiance becomes more noticeable, too.

\section{Superradiant Instability with Lorentz-Violating parameter}\label{sec:SI}
Above, we have discussed on the superradiance amplification of a bosonic field such as
massive scalar field due to scattering of a bumblebee-Kerr black hole. Given the fact
that superradiance scattering does not necessarily mean instability so here we extend it
within a related issue knows as \textit{``black hole bomb''} for seeing the role of Lorentz-violating
parameter on superradiant instability. Technically speaking, the instability of the composite
system of Kerr background and massive scalar perturbations may be due to the massive modes which are
captured inside the effective potential well outside the black hole. These massive modes,
in essence, behave like a mirror so that by returning the reflection waves toward black hole and subsequently
their amplifying by creating bounce back and forth moves, give rise to an superradiantly instability known as
black hole bomb. Overall, to trigger this instability, the existence of a trapping
potential well outside the black hole is crucial in addition to the existence of an ergo-region, to happening
the superradiant amplification. In the following, we will investigate this for a Kerr-like background include
Lorentz symmetry breaking characterized by constant parameter $\alpha$ which is subjecting to massive scalar perturbation.

Beginning the radial Klein-Gordon equation \eqref{eq:ODE_rad}, we have
\begin{equation}\label{eq:kg}
\Delta_{\alpha}{{d}\over{dr}}\Big(\Delta_{\alpha}{{dF_{\omega lm}}\over{dr}}\Big)+\xi F_{\omega lm}=0~,
\end{equation}
where in the slowly rotating regime ($\hat{a}\omega\ll1$)
\begin{equation}\label{eq:M}
\xi\equiv \bigg((r^2+\hat{a}^2)\omega-m\hat{a}\bigg)^2+\Delta_\alpha\bigg(2m\hat{a}\omega-l(l+1)-\mu^2(r^2+\hat{a}^2)\bigg)~.
\end{equation}
Demanding the black hole bomb mechanism, we should have the following solutions
for the radial equation (\ref{eq:kg})
\begin{eqnarray}\label{eq:so}
F_{\omega lm}\sim\left\{
\begin{array}{ll}
e^{-i (\omega-m\hat{\Omega})r_*}\ \ \text{ as }\ r\rightarrow r_{eh}\ \
(r_*\rightarrow -\infty)  \\\\
\frac{e^{-\sqrt{\mu^2-\omega^2}r_*}}{r}\ \ \text{ as }\
r\rightarrow\infty\ \ \ \ (r_*\rightarrow \infty)
\end{array}
\right.
\end{eqnarray}
The solutions above, are represent this physical boundary conditions that the scalar wave
at the black hole horizon is purely ingoing while at spatial infinity it is a decaying exponentially
(bounded) solution, provided that $\omega^2<\mu^2$.

By regarding a new radial function as follows
\begin{equation}\label{eq:n}
\psi_{\omega lm}\equiv \sqrt{\Delta_\alpha}F_{\omega lm}\  ,
\end{equation}
in the radial equation (\ref{eq:kg}), one come to
\begin{equation}\label{eq:RW}
\bigg({{d^2}\over{dr^2}}+\omega^2-V\bigg)\psi_{\omega lm}=0\  ,
\end{equation}
with
\begin{equation}\label{eq:RW2}
\omega^2-V={{\xi+M^2-\hat{a}^2}\over{\Delta_\alpha^2}}\ .
\end{equation}
where is indeed Regge-Wheel equation. With a straightforward calculation, one can show that
the asymptotic form of the effective potential $V$, by discarding terms $\mathcal{O}(1/r^2)$
is in the form below
\begin{equation}\label{eq:V}
V(r)=\mu^2-\frac{4M\omega^2}{r}+(\alpha+1) \frac{2M\mu^2}{r}~.
\end{equation}
\begin{figure}[!ht]
	\begin{center}
			\includegraphics[scale=0.35]{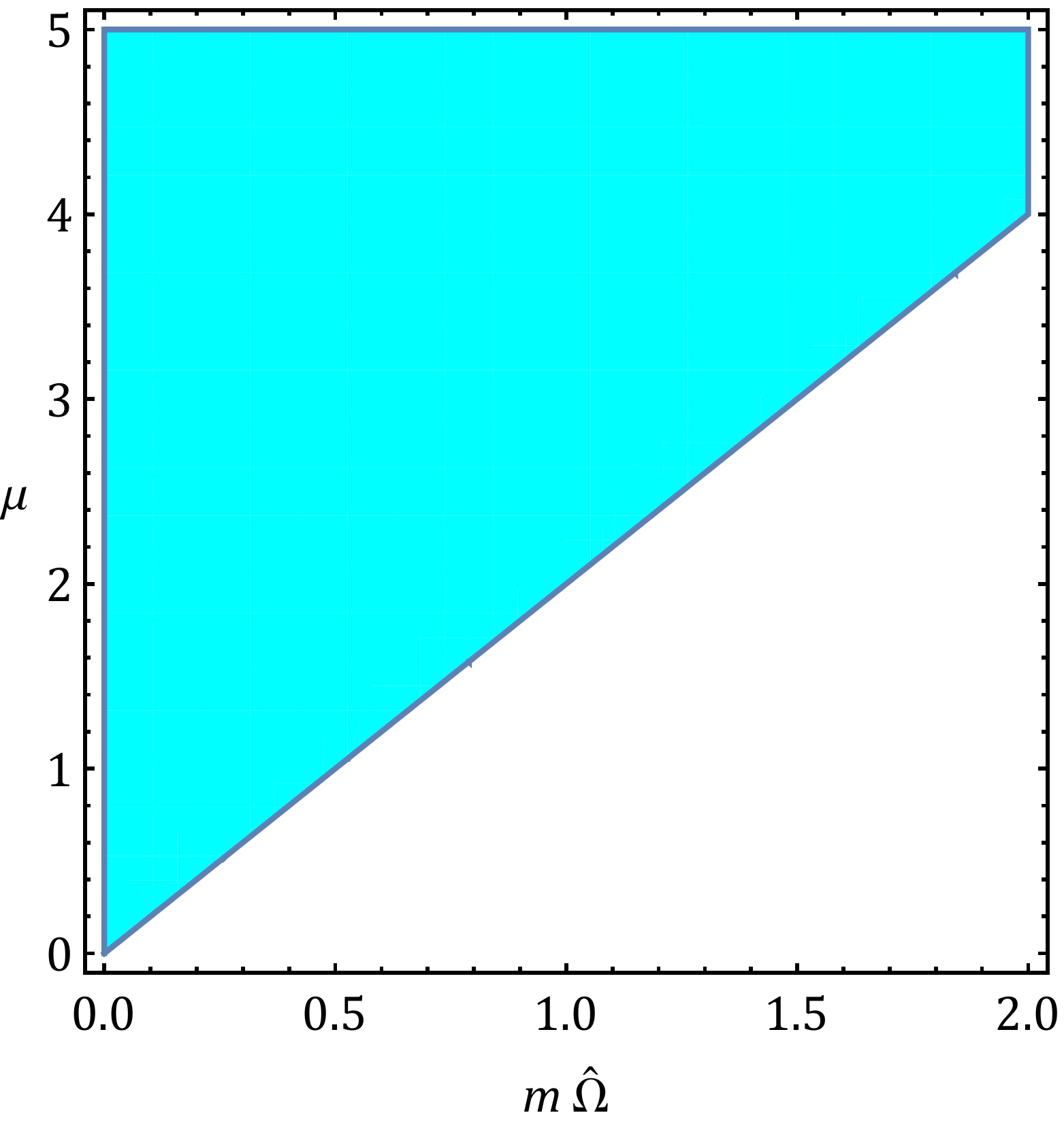}
			\includegraphics[scale=0.35]{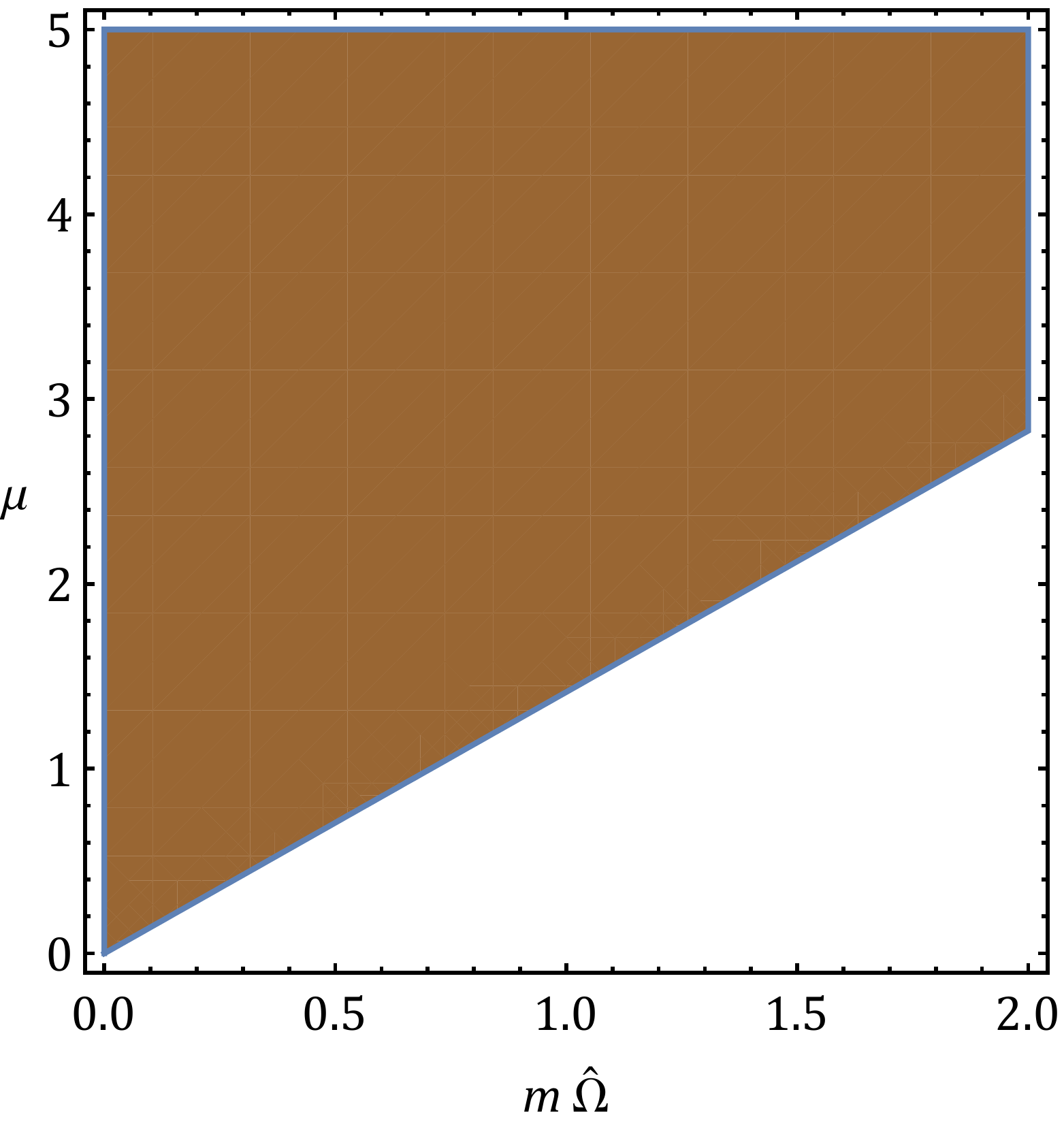}
            \includegraphics[scale=0.35]{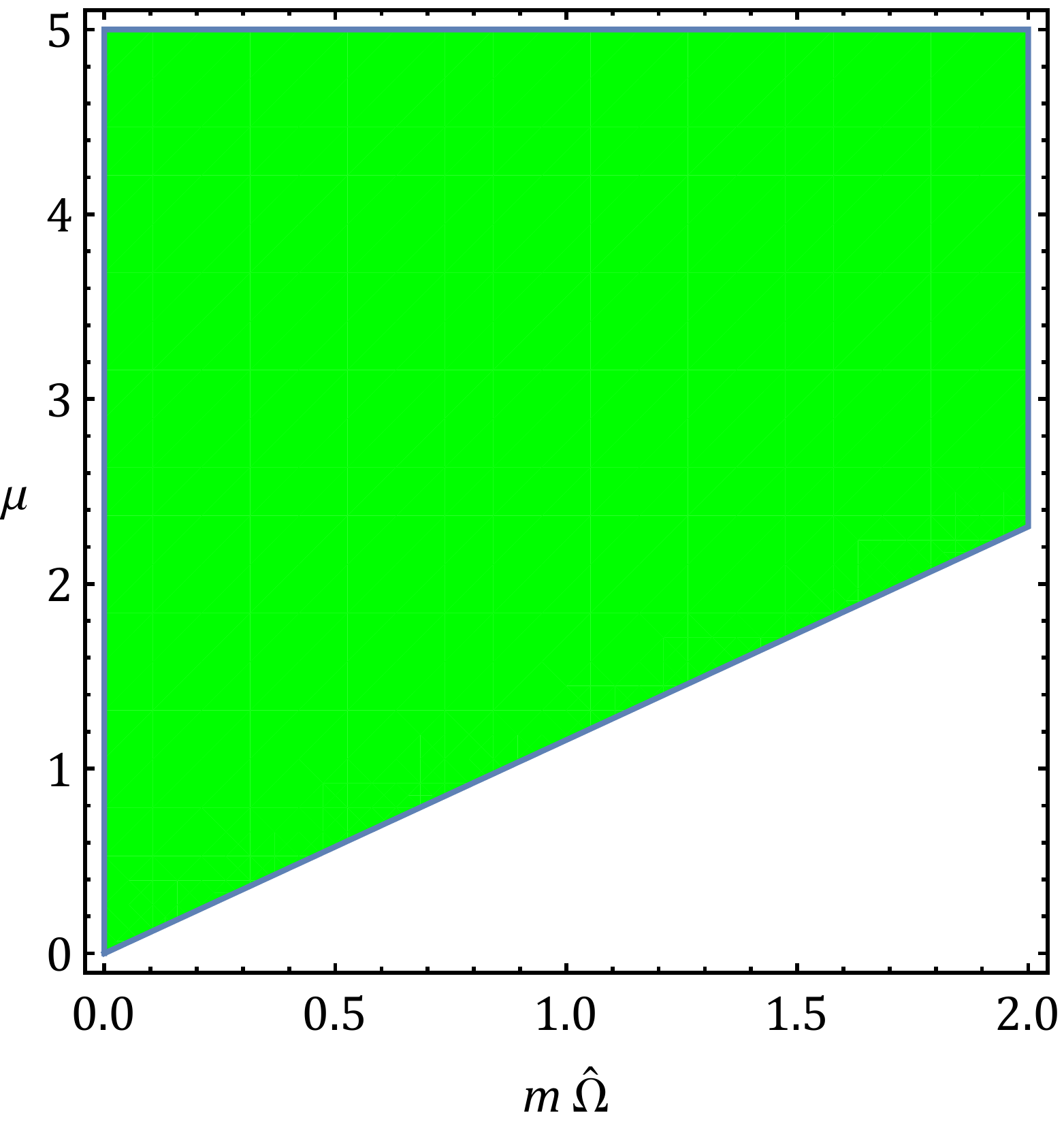}
             \caption{The parameter space $(m\hat{\Omega},\mu)$ of massive scalar
              field in the background of bumblebee-Kerr spacetime include Lorentz violating parameter $\alpha$
               with different values:$ -0.5,~ 0,~ 0.5$ from left to right, respectively. Colored and non-colored
               areas respectively represent regions with stable and unstable dynamics for the massive scalar field.}
		\label{Z4}
	\end{center}
     \end{figure}

To releasing a trapping well by above effective potential it is necessary that its asymptotic
derivative be positive i.e $V'\to 0^+$ as $r\to\infty$ \cite{Hod:2012zza}. As a result, one
acquire the following modified regime
\begin{equation}\label{eq:reg}
\frac{\alpha+1}{2}\mu^2<\omega^2<\mu^2\ .
\end{equation} in which the bound states may capture. It is very obvious that
the above regime is well-defined provided that $\alpha>-1$.
We have shown before that the superradiance condition for amplification of the
scattered waves from underlying background is $\omega<m\hat{\Omega}$ where in case of facing
with modified instability regime (\ref{eq:reg}), one finally realizes following restricted regime
\begin{equation}
\sqrt{\frac{\alpha+1}{2}}\mu<\omega<m\hat{\Omega }~,
\end{equation} in which the integrated system of Kerr bumblebee black hole and massive scalar
field, may experience a superradiant instability known as black hole bomb. Namely, the dynamics
of the massive scalar field in confronting with a Lorentz symmetry violating background,
is expected to remain stable in the complementary regime $\mu\geq\sqrt{\frac{2}{\alpha+1}} m\hat{\Omega}$. As can be seen from Fig. (\ref{Z4}), by going the value of
$\alpha$ from negative to positive ones within the allowed range $\alpha>-1$, then the superradiant
instability regions become smaller. In other words, as can be seen, the massive scalar filed within a Kerr like background with the positive values of $\alpha$ has a lower chance of causing instability relative to their negative counterparts.
This is interesting in the sense that in case of $\alpha<0$ as the massive scalar wave superradiance enhances in compared to its standard counterpart, it also has a better chance of making instability, and vice versa for $\alpha>0$. However, it is important to remember that there is no requirement that
in case of larger/smaller energy extraction from background we necessarily deal with a stronger/weaker superradiant instability \footnote{A similar situation has already been seen in a composite system of charged massive scalar field
and RN black hole so that in the regime $Q\leq2\sqrt{2}/3M$, the black hole experiences superradiance without instability,
see \cite{Hod:2013eea} for more details.}.

Recently in \cite{Creci:2020mfg} by employing a numerical method claimed that the superradiant instability
due to competing effects arose from the mass losing and spin-down can lead to a decrease and an increase
of the shadow radius, respectively. So, dependent on dominating each of these two effects, then it can be said
that the superradiant instability decreases or increases the size of the shadow, respectively. Besides,
as before mentioned, we know from \cite{Ding:2019mal} that moving from $\alpha<0$ to $\alpha>0$ , leads to a decrease of shadow radius.
With these descriptions, it seems that for $\alpha<0$, the effect of spin-down is dominates on the mass loss
while for $\alpha>0$, its opposite happens.
\section{Conclusion}\label{sec:Con}
In this paper, we have studied the superradiance phenomenon and its related instability for a spinning-like black hole in the Einstein-bumblebee modified gravity. The key property of such a theory of gravity actually is  the presence of a vector field known as bumblebee field with non-zero vacuum expectation value, representing spontaneously breaking of Lorentz symmetry. With the ask of how
the spontaneous Lorentz symmetry breaking affects the superradiance scattering and corresponding instability, we have considered a composited system of the massive scalar perturbation and a small rotating black hole admitted by the theory of gravity at hand.
By investigating the low-frequency limit through employing asymptotic matching method,  we analytical shown that the massive scalar wave amplifies with the Lorentz breaking constant $\alpha<0$, and weakens with $\alpha>0$.
Next, from the analytical study of superradiant instability via addressing the issue of the black hole bomb, we found that the Lorentz breaking constant affects the instability regime so that the background with $\alpha<0$ has more chance of unstable dynamics for the scalar field while with $\alpha>0$ less. As a result, the Lorentz breaking constant affects identically the superradiance scattering and relevant instability.
Finally, with the integration of our results with the discussions released
in Refs. \cite{Ding:2019mal} and \cite{Creci:2020mfg}, we have argued that in the superradiant scattering with $\alpha<0$ and $\alpha>0$, respectively
spin-down and mass loss of black hole (as two competing influences at play) separately become dominate.
By imposing observational constraints on the Lorentz breaking constant $\alpha$ in the future,
it may be ruled out one of the two possibilities above.

\vspace{1cm}
{\bf Acknowledgments:}
I would like to thank Carlos Herdeiro for fruitful comments and discussions on the manuscript.
I thank Sunny Vagnozzi for the detailed discussions on bumblebee gravity.
I also need to appreciate Soroush Shakeri for his help in final editing the manuscript.

\end{document}